\documentclass[journal,onecolumn,draftcls,12pt]{IEEEtran}
\usepackage[T1]{fontenc}
\pdfoutput=1

\makeatletter
\def\endthebibliography{%
	\def\@noitemerr{\@latex@warning{Empty `thebibliography' environment}}%
	\endlist
}
\makeatother

\IEEEoverridecommandlockouts
\usepackage{cite}
\usepackage{amsmath,amssymb,amsfonts}
\usepackage{algorithm}
\usepackage{algorithmic}
\usepackage{etoolbox}  
\usepackage{bbm}
\usepackage{amsmath}
\usepackage{tabularx}
\usepackage{amssymb}
\usepackage{threeparttable}
\usepackage{booktabs}
\usepackage{amsthm}
\usepackage{mathtools}
\usepackage{bm}
\makeatletter
\patchcmd{\algorithmic}{\addtolength{\ALC@tlm}{\leftmargin} }{\addtolength{\ALC@tlm}{\leftmargin}}{}{}
\makeatother

\makeatletter
\newcommand\fs@betterruled{%
	\def\@fs@cfont{\bfseries}\let\@fs@capt\floatc@ruled
	\def\@fs@pre{\vspace*{5pt}\hrule height.8pt depth0pt \kern2pt}%
	\def\@fs@post{\kern2pt\hrule\relax}%
	\def\@fs@mid{\kern2pt\hrule\kern2pt}%
	\let\@fs@iftopcapt\iftrue}
\floatstyle{betterruled}
\restylefloat{algorithm}
\makeatother

\usepackage{tikz}
    \usetikzlibrary{shapes.arrows}

\usepackage{pgfplots}
\usepackage{adjustbox}

\usepackage{gincltex}
\usepgfplotslibrary{fillbetween}
\usetikzlibrary{plotmarks}
\usetikzlibrary{patterns}
\usepackage[aboveskip=-5pt]{subcaption}
\usepackage{graphicx}
\usepackage{textcomp}
\usepackage{xcolor}
\usepackage{authblk}
\def\BibTeX{{\rm B\kern-.05em{\sc i\kern-.025em b}\kern-.08em
		T\kern-.1667em\lower.7ex\hbox{E}\kern-.125emX}}
		
\pgfplotsset{compat=1.14}
		
\usepackage{glossaries}

\newacronym{opf}{OPF}{Oldest Packet First}
\newacronym{isl}{ISL}{Inter-Satellite Link}
\newacronym{jfi}{JFI}{Jain Fairness Index}
\newacronym{aoi}{AoI}{Age of Information}
\newacronym{paoi}{PAoI}{Peak Age of Information}
\newacronym{pdf}{PDF}{Probability Density Function}
\newacronym{cdf}{CDF}{Cumulative Density Function}
\newacronym{fcfs}{FCFS}{First Come First Serve}
\newacronym{ar}{AR}{Augmented Reality}
\newacronym{vr}{VR}{Virtual Reality}
\newacronym{qoe}{QoE}{Quality of Experience}
\newacronym{lcfs}{LCFS}{Last Come First Serve}
\newacronym{iot}{IoT}{Internet of Things}
\newacronym{pec}{PEC}{Packet Erasure Channel}
\newacronym[plural=MDPs,firstplural=Markov Decision Processes (MDPs)]{mdp}{MDP}{Markov Decision Process}
\newacronym{mdc}{MDC}{Multiple Description Coding}
\newacronym{fhw}{FHW}{Flatto-Hahn-Wright}
\newacronym[plural=RATs,firstplural=Radio Access Technologies (RATs)]{rat}{RAT}{Radio Access Technology}
\newacronym{mptcp}{MPTCP}{Multipath TCP}
\newacronym{sctp}{SCTP}{Stream Control Transmission Protocol}
\newacronym{leap}{LEAP}{Latency-controlled End-to-End Aggregation Protocol}
\newacronym{rtp}{RTP}{Real-time Transport Protocol}
\newacronym{mpmtp}{MPMTP}{Multipath Multimedia Transport Protocol}

\definecolor{violet}{rgb}{0.6,0,0.6}%
\definecolor{orange_D}{rgb}{1,0.3,0}%
\definecolor{cyan}{rgb}{0,0.67,0.64}%
\definecolor{red}{rgb}{0.9,0,0}%
\definecolor{green}{rgb}{0,0.8,0}%
\definecolor{yellow}{rgb}{1,0.8,0}

\def \fwidth{0.6\columnwidth}
\def \fheight {0.2\columnwidth}

\def \sfwidth{0.8\linewidth}
\def \sfheight {0.4\linewidth}

\begin{document}

\title{Latency and Information Freshness in Multipath Communications for Virtual Reality}

\author{Federico Chiariotti, Beatriz Soret, Petar Popovski}

\affil{
Department of Electronic Systems, Aalborg University, Denmark\\
Email:{\texttt{\{fchi,bsa,petarp\}@es.aau.dk}}} 

\maketitle


\begin{abstract}
Wireless \gls{vr} and \gls{ar} will contribute to people increasingly working and socializing remotely. However, the \gls{vr}/\gls{ar} experience is very susceptible to various delays and timing discrepancies, which can lead to motion sickness and discomfort. This paper models and exploits the existence of multiple paths and redundancy to improve the timing performance of wireless \gls{vr} communications. We consider \gls{mdc}, a scheme where the video stream is encoded in $\mathcal{Q}$ streams ($\mathcal{Q}=2$ in this paper) known as descriptors and delivered independently over multiple paths. We also consider an alternating scheme, that simply switches between the paths. We analyze the full distribution of two relevant metrics: the packet delay and the \gls{paoi}, which measures the freshness of the information at the receiver. The results show interesting trade-offs between picture quality, frame rate, and latency: full duplication results in fewer lost frames, but a higher latency than schemes with less redundancy. Even the simple alternating scheme can outperform duplication in terms of \gls{paoi}, but \gls{mdc} can exploit the independent decodability of the descriptors to deliver a basic version of the frames faster, while still getting the full-quality frames with a slightly higher delay.
\end{abstract}

\IEEEpeerreviewmaketitle
\glsresetall

\section{Introduction}

Over the past decade, \gls{vr} and \gls{ar} have exited the realm of science fiction and become an increasingly commonplace reality in a variety of fields, from education~\cite{kavanagh2017systematic} and medicine~\cite{freeman2017virtual} to tourism and industry~\cite{farshid2018go}. The worldwide COVID-19 pandemic has only accelerated the development of these technologies~\cite{singh2020significant}, as the need for social distancing has moved several aspects of work and personal life online. \gls{vr} is also crucial to the \emph{digital twin} concept, which enables remote inspection and operation of cyber-physical systems, and for human-in-the-loop control, in which parts of a manufacturing system are automated, while other parts are controlled directly by a human operator~\cite{chakraborti2018projection,broring2021intelliot}. In these scenarios, the timeliness of the operation is crucial, as delays can have a significant impact on the performance and even safety of the machinery. Moreover, the necessity of communication cables severely limits the mobility of both the operators and the autonomous robots.

Computational offloading can enable smooth operation on smaller devices or with mobile nodes~\cite{braud2020multipath}, freeing \gls{ar} and \gls{vr} users from being tied to desktop-level computing power through direct wires which can impede movement and limit immersiveness.
However, these applications can suffer from significant timeliness-related issues on wireless networks~\cite{li2015influence}: as the sense of presence and immersion is critical, and the size of each omnidirectional frame can be huge, this puts a strain on the notoriusly volatile wireless links, with a high risk of congestion and sudden delay increases, which are perceived by the user as an annoying loss of smoothness in the \gls{vr} experience. Adaptive video content can reduce delay by compressing the video, trading video quality for a smaller, more predictable delay~\cite{wu2016trading}, but the unpredictability of the wireless propagation environment can make this task very complex.

In this context, there are two metrics that can be used to evaluate the timeliness of the communication: the first, and most traditional, is the latency of the transmission, i.e., the time between when a frame is generated at the source and when it is delivered to the receiver and displayed to the user, while the second is the \gls{aoi}. \gls{aoi} is a metric that has attracted a significant interest in the research community since its inception in the early 2010s~\cite{kaul2012real}, as it can better represent the delay perceived by users in real-time applications. The age does not just measure the time between when a frame is generated and when it is delivered, but keeps increasing until the next frame has been delivered. As the name suggests, the \gls{aoi} is the \emph{age} of the frame that is currently on the \gls{vr} display. Intuitively, \gls{aoi} is more useful for control tasks, as the age of the information shown to the user will affect both the smoothness of the \gls{vr} experience and the control performance: if the user makes a command, it will be based on old information, and the older the scene that the user can see, the less immediate that command will get. In particular, we consider the \gls{paoi}, which measures the age right before the next frame arrives, which is a good proxy for the worst-case timing discrepancy of a virtual reaction.

In this work, we consider a theoretical model of multipath \gls{vr} transmission, deriving the complete distribution of the latency and \gls{paoi} analytically with four different transmission schemes. In particular, we consider both schemes that try to reduce the load on the multipath connection by splitting the traffic between the two paths and a scheme that replicates the frames on both paths to protect them from errors and delays of the individual paths. \gls{mdc}~\cite{kazemi2014review} is a scheme that can strike a balance between smoothness and quality, as it encodes the frame into multiple \emph{descriptors}, each of which can then be delivered independently. Each descriptor can be decoded individually, resulting in a lower-quality representation, but larger numbers of descriptors can be combined to recover higher-quality representations if they are delivered on time. We draw some considerations on the performance of each scheme, finding interesting results in the trade-off between reliability (i.e., delivering as many frames as possible in error-prone scenarios), frame rate and quality, and latency or age.

The rest of the paper is organized as follows: Sec.~\ref{sec:related} presents an overview of the related work on multipath multimedia communications, queuing models and \gls{aoi}. Our overall system model is described in Sec.~\ref{sec:system}, and the analyses for the alternating and coded systems are described in Sec.~\ref{sec:an_alt} and Sec.~\ref{sec:an_mult}, respectively. Sec.~\ref{sec:sim} then presents our simultion settings and results, and Sec.~\ref{sec:concl} concludes the paper and presents some avenues of future work.

\section{Related work}\label{sec:related}

\subsection{Multipath multimedia communications}

Multipath communications can be exploited for very different applications and purposes, such as increasing the network capacity~\cite{raiciu2011improving}, increasing the reliability~\cite{rao2018packet}, or reducing the latency~\cite{chiariotti2019analysis}.
Path diversity is often used as a means to apply other redundancy schemes, such as error control coding~\cite{ferlin2018mptcp}.
In the case of multimedia content in real time, the use of multiple paths to overcome possible failures and blockages on any single link has been the subject of intense investigation over the past few years. For \gls{ar}/\gls{vr}, the higher bitrate required to transmit 360$^\circ$ frames can exacerbate these issues, making the benefits of multipath delivery even more evident: a higher capacity and resilience to failures or capacity drops on any single path can significantly improve the \gls{qoe} and the smoothness of remote control operations.

The \gls{iot} has seen a significant expansion in the past decade, and previously impossible video sensor applications are now commonplace: in this case, real-time operation can be required to monitor a process or an area remotely, or to control a robot's activity. In this field, energy-efficiency is a significant concern~\cite{mekonnen2017energy}, and communication strategies need to take it into account: a recent survey on multipath routing in this context~\cite{hasan2017survey} describes several techniques that have been proposed by the research community. In this case, video quality is generally low, and the most pressing concern is routing rather than bitrate adaptation. Interestingly, the same type of concepts can also apply to \gls{vr} services in heterogeneous 5G networks~\cite{ge2017multipath}: latency and energy-efficiency are critical metrics in small cell networks.

There has also been significant activity on the transport layer to enable end-to-end multipath communications. As the bottlenecks in the paths might not be directly observable, this is a significantly more complex problem, which intersects congestion control research and highlights some of the issues of the currently used transport protocols. The \gls{mptcp} standard~\cite{rfc6824} was developed as an extension of TCP to multipath connections, enabling faster and more reliable delivery, but it has been shown to be unsuitable for real-time media. While early studies~\cite{brosh2010delay} had promising outcomes, \gls{mptcp} fell prey to the same basic issues that standard TCP has with real-time traffic~\cite{chen2013measurement}: namely, the potentially long delays caused by the congestion control and retransmission mechanisms if the capacity drops. Furthermore, scheduling packets over the available paths is another problem specific to multipath transmission~\cite{hurtig2018low}, which can compound with the congestion issues and interact with congestion control in non-obvious ways~\cite{khalili2013mptcp}. As it is the most common protocol, however, \gls{mptcp} is used in a number of \gls{vr} solutions~\cite{silva2020avira}, which devise application-level schemes to work around its issues.

In order to overcome the limitations inherent in \gls{mptcp}, there is active research on multipath versions of other protocols such as the \gls{rtp}~\cite{singh2013mprtp} and the \gls{sctp}~\cite{wallace2014concurrent}, but the lack of widespread adoption makes these protocols less likely candidates for future adoption. More recently, other solutions relying on entirely new protocols have been developed: one example is the \gls{mpmtp}~\cite{kwon2014mpmtp}, which uses Raptor codes to recover from packet losses without retransmissions. The more recent \gls{leap} protocol~\cite{chiariotti2019analysis} also uses coding, working on blocks with adaptable size and providing reliable latency guarantees. For a more comprehensive study of multipath transport layer solutions, we refer the reader to~\cite{polese2019survey}. In most of these applications, the basic trade-off is between meeting the application's latency, throughput, and reliability requirements and reducing the impact on other flows, as using resources on multiple paths is a greedy option that should be limited to necessary uses.

\subsection{Queuing models} \label{sec:fork}
Our system model is based on a fork-join queue~\cite{kim1989fork}, where incoming tasks are split into several servers and joined again before departing the system. This model has been used for all kinds of parallel multitasking in computation and communications networks~\cite{khudabukhsh2017optimizing}. Most of the works assume that, at any time instant, tasks can be canceled and abandon their respective queue.
\cite{joshi2017cloud} studies path redundancy in the context of cloud systems with a fork-join model, to understand the trade-off latency-computing cost.  \cite{shah2016redundant} analyzes the transmission of redundant requests to multiple servers for a faster execution in terms of average latency, at the cost of increased system load. It is observed that not having redundancy is optimal for highly loaded systems if service times are memoryless. The authors in \cite{Sun2016OnDS} present a study on delay-optimal scheduling of replications in centralized and distributed multi-server systems. 

There has also been some work on bounds to worst-case performance in fork-join queues, mostly concerned with the tail of the latency distribution~\cite{rizk2016stochastic}. These delay bounds concern themselves with Markovian arrival processes~\cite{fidler2016non}, as they model parallel or distributed processing of randomly arriving data. To the best of our knowledge, the question of delay bounds with periodic traffic, such as video or \gls{vr}/\gls{ar} frames, has not been investigated in either the communication or distributed processing literature.

\subsection{Age of Information}

Since the concept of \gls{aoi} was introduced in the early 2010s~\cite{kaul2012real}, it has been the subject of intense study, being the preferred latency metric for real-time video streaming and context-aware \gls{iot} applications \cite{abdelmagid2019ontheroleofaoi}. In these applications, the end receiver is interested in a fresh knowledge of the remotely controlled system, rather than the packet delay. Most theoretical results derive the average \gls{aoi} or the \gls{paoi}. Much smaller is the number of works deriving higher moments or the full distribution of the age (see e.g. ~\cite{chiariotti2019analysis}\cite{inoue2019general}), although a reliable system design requires knowing the probability of occurrence of rare, but extremely damaging events.

Initial studies focused on simple queuing systems with a single node and unicast scheduled transmissions~\cite{kosta2017age,yates2020tutorial}. However, there is already a significant number of papers addressing more complex communication scenarios and topologies, like for instance models for random access and ALOHA~\cite{chen2021rach}\cite{munari2021irsa}, multi-cast and broadcast~\cite{li2020age}, or multi-hop transmissions~\cite{chiariotti2020peak,akar2020finding}.

None of the fork-join queuing works listed in Section~\ref{sec:fork} analyzes the \gls{aoi}. To the best of our knowledge, the only exceptions are \cite{buyukates2020timely}, which gives the average age for distributed computing with $k$ over $n$ coding, and \cite{talak2021age}, which addresses the age-delay trade-off considering the scheduler and routing (selected path) of an M-server system with Poisson arrivals. They prove that a system designed to minimize the age will do it at the expenses of high waiting times and service times for the packets that do not contribute to the age metric (the \emph{non-informative} packets) and the average and variance of the packet delay will therefore increase enormously. 

Unlike most of the fork-related works, we do not consider removal of task in the buffers, since age-sensitive applications will typically have no feedback and there is no way for the transmitter to know when the replicas/pieces of the frame have arrived to any of the receivers. 

\section{System model}\label{sec:system}

%

\begin{figure}
    \centering
	\begin{subfigure}[b]{.49\linewidth}
	    \centering
        \vspace{0.3cm}
\begin{tikzpicture}[>=latex]

\draw [fill=white!80!green,draw=none](-1.5,1.05) rectangle (-1.8,0.45); 
\draw [fill=white!80!blue,draw=none] (-1.8,1.05) rectangle (-2.1,0.45); 
\draw [fill=white!80!red,draw=none] (-2.1,1.05) rectangle (-2.4,0.45); 
\draw [fill=white!80!orange,draw=none] (-2.4,1.05) rectangle (-2.7,0.45); 

\node at (-1.65,0.75) {\small 1};
\node at (-1.95,0.75) {\small 2};
\node at (-2.25,0.75) {\small 3};
\node at (-2.55,0.75) {\small 4};

\draw [fill=white!80!green,draw=none] (1.6,1.8) rectangle (1.3,1.2); 
\draw [fill=white!90!blue,draw=none] (1,0.3) rectangle (1.3,-0.3);

\node at (1.45,1.5) {\small 1};
\node at (1.15,0) {\small 2};
  
\draw (-3.1,1.05) -- ++(1.6cm,0) -- ++(0,-0.6cm) -- ++(-1.6cm,0);
\foreach \i in {1,...,4}
  \draw (-1.5cm-\i*0.3cm,1.05) -- +(0,-0.6cm);
  
\draw (0,0.3) -- ++(1.6cm,0) -- ++(0,-0.6cm) -- ++(-1.6cm,0);
\foreach \i in {1,...,4}
  \draw (1.6cm-\i*0.3cm,0.3) -- +(0,-0.6cm);
  
\draw (0,1.8) -- ++(1.6cm,0) -- ++(0,-0.6cm) -- ++(-1.6cm,0);
\foreach \i in {1,...,4}
  \draw (1.6cm-\i*0.3cm,1.8) -- +(0,-0.6cm);

\draw (1.9,0cm) circle [radius=0.3cm];
\draw (1.9,1.5cm) circle [radius=0.3cm];

\draw[->,dashed] (-1.5,0.75) -- +(1.5cm,0.75cm);
\draw[->,dashed] (-1.5,0.75) -- +(1.5cm,-0.75cm);

\draw[<->]  (-1.3,1.1)  to[out=-30,in=30] (-1.3,0.4);

\node at (1.9,0) {\small $\mu_2$};
\node at (1.9,1.5) {\small $\mu_1$};

\draw[->] (2.2,0) -- +(1cm,0);
\draw[->] (2.2,1.5) -- +(1cm,0);

\end{tikzpicture}
\vspace{0.3cm}
        \caption{Alternating transmission.}
        \label{fig:alt}
    \end{subfigure}	
	\centering
	\begin{subfigure}[b]{.49\linewidth}
	    \centering
        \vspace{0.3cm}
\begin{tikzpicture}[>=latex]

\draw [fill=white!80!green,draw=none](-1.5,1.05) rectangle (-1.8,0.45); 
\draw [fill=white!80!blue,draw=none] (-1.8,1.05) rectangle (-2.1,0.45); 
\draw [fill=white!80!red,draw=none] (-2.1,1.05) rectangle (-2.4,0.45); 
\draw [fill=white!80!orange,draw=none] (-2.4,1.05) rectangle (-2.7,0.45); 

\node at (-1.65,0.75) {\small 1};
\node at (-1.95,0.75) {\small 2};
\node at (-2.25,0.75) {\small 3};
\node at (-2.55,0.75) {\small 4};

\draw [fill=white!80!green,draw=none] (1.6,1.8) rectangle (1.3,1.2); 
\draw [fill=white!80!green,draw=none] (1.6,0.3) rectangle (1.3,-0.3);

\node at (1.45,1.5) {\small 1};
\node at (1.45,0) {\small 1};
  
\draw (-3.1,1.05) -- ++(1.6cm,0) -- ++(0,-0.6cm) -- ++(-1.6cm,0);
\foreach \i in {1,...,4}
  \draw (-1.5cm-\i*0.3cm,1.05) -- +(0,-0.6cm);
  
\draw (0,0.3) -- ++(1.6cm,0) -- ++(0,-0.6cm) -- ++(-1.6cm,0);
\foreach \i in {1,...,4}
  \draw (1.6cm-\i*0.3cm,0.3) -- +(0,-0.6cm);
  
\draw (0,1.8) -- ++(1.6cm,0) -- ++(0,-0.6cm) -- ++(-1.6cm,0);
\foreach \i in {1,...,4}
  \draw (1.6cm-\i*0.3cm,1.8) -- +(0,-0.6cm);

\draw (1.9,0cm) circle [radius=0.3cm];
\draw (1.9,1.5cm) circle [radius=0.3cm];

\draw[->] (-1.5,0.75) -- +(1.5cm,0.75cm);
\draw[->] (-1.5,0.75) -- +(1.5cm,-0.75cm);

\node at (1.9,0) {\small $\mu_2$};
\node at (1.9,1.5) {\small $\mu_1$};

\draw[->] (2.2,0) -- +(1cm,0);
\draw[->] (2.2,1.5) -- +(1cm,0);

\end{tikzpicture}
\vspace{0.3cm}
        \caption{Replicated transmission.}
        \label{fig:rep}
    \end{subfigure}	
	\begin{subfigure}[b]{.49\linewidth}
	    \centering
        \vspace{0.3cm}
\begin{tikzpicture}[>=latex]

\draw [fill=white!80!green,draw=none](-1.5,1.05) rectangle (-1.8,0.45); 
\draw [fill=white!80!blue,draw=none] (-1.8,1.05) rectangle (-2.1,0.45); 
\draw [fill=white!80!red,draw=none] (-2.1,1.05) rectangle (-2.4,0.45); 
\draw [fill=white!80!orange,draw=none] (-2.4,1.05) rectangle (-2.7,0.45); 

\node at (-1.65,0.75) {\small 1};
\node at (-1.95,0.75) {\small 2};
\node at (-2.25,0.75) {\small 3};
\node at (-2.55,0.75) {\small 4};

\draw [fill=white!90!green,draw=none] (1.6,1.8)--(1.3,1.8)--(1.3,1.2)--cycle;
\draw [fill=white!90!green,draw=none] (1.6,0.3)--(1.6,-0.3)--(1.3,-0.3)--cycle; 

\draw[-,dashed] (1.6,1.8) -- (1.3,1.2);
\draw[-,dashed] (1.6,0.3) -- (1.3,-0.3);

\node at (1.45,1.5) {\small 1};
\node at (1.45,0) {\small 1};
  
\draw (-3.1,1.05) -- ++(1.6cm,0) -- ++(0,-0.6cm) -- ++(-1.6cm,0);
\foreach \i in {1,...,4}
  \draw (-1.5cm-\i*0.3cm,1.05) -- +(0,-0.6cm);
  
\draw (0,0.3) -- ++(1.6cm,0) -- ++(0,-0.6cm) -- ++(-1.6cm,0);
\foreach \i in {1,...,4}
  \draw (1.6cm-\i*0.3cm,0.3) -- +(0,-0.6cm);
  
\draw (0,1.8) -- ++(1.6cm,0) -- ++(0,-0.6cm) -- ++(-1.6cm,0);
\foreach \i in {1,...,4}
  \draw (1.6cm-\i*0.3cm,1.8) -- +(0,-0.6cm);

\draw (1.9,0cm) circle [radius=0.3cm];
\draw (1.9,1.5cm) circle [radius=0.3cm];

\draw[->] (-1.5,0.75) -- +(1.5cm,0.75cm);
\draw[->] (-1.5,0.75) -- +(1.5cm,-0.75cm);

\node at (1.9,0) {\small $\mu_2$};
\node at (1.9,1.5) {\small $\mu_1$};

\draw[->] (2.2,0) -- +(1cm,0);
\draw[->] (2.2,1.5) -- +(1cm,0);

\end{tikzpicture}
\vspace{0.3cm}
        \caption{Split transmission.}
        \label{fig:split}
    \end{subfigure}	
        	\begin{subfigure}[b]{.49\linewidth}
	    \centering
        \vspace{0.3cm}
\begin{tikzpicture}[>=latex]

\draw [fill=white!80!green,draw=none](-1.5,1.05) rectangle (-1.8,0.45); 
\draw [fill=white!80!blue,draw=none] (-1.8,1.05) rectangle (-2.1,0.45); 
\draw [fill=white!80!red,draw=none] (-2.1,1.05) rectangle (-2.4,0.45); 
\draw [fill=white!80!orange,draw=none] (-2.4,1.05) rectangle (-2.7,0.45); 

\node at (-1.65,0.75) {\small 1};
\node at (-1.95,0.75) {\small 2};
\node at (-2.25,0.75) {\small 3};
\node at (-2.55,0.75) {\small 4};

\draw [fill=white!90!green,draw=none] (1.6,1.2) rectangle (1.3,1.65); 
\draw [fill=white!90!green,draw=none] (1.6,0.15) rectangle (1.3,-0.3);

\draw[-,dashed] (1.6,1.65) -- (1.3,1.65);
\draw[-,dashed] (1.6,0.15) -- (1.3,0.15);

\node at (1.45,1.5) {\small 1};
\node at (1.45,0) {\small 1};
  
\draw (-3.1,1.05) -- ++(1.6cm,0) -- ++(0,-0.6cm) -- ++(-1.6cm,0);
\foreach \i in {1,...,4}
  \draw (-1.5cm-\i*0.3cm,1.05) -- +(0,-0.6cm);
  
\draw (0,0.3) -- ++(1.6cm,0) -- ++(0,-0.6cm) -- ++(-1.6cm,0);
\foreach \i in {1,...,4}
  \draw (1.6cm-\i*0.3cm,0.3) -- +(0,-0.6cm);
  
\draw (0,1.8) -- ++(1.6cm,0) -- ++(0,-0.6cm) -- ++(-1.6cm,0);
\foreach \i in {1,...,4}
  \draw (1.6cm-\i*0.3cm,1.8) -- +(0,-0.6cm);

\draw (1.9,0cm) circle [radius=0.3cm];
\draw (1.9,1.5cm) circle [radius=0.3cm];

\draw[->] (-1.5,0.75) -- +(1.5cm,0.75cm);
\draw[->] (-1.5,0.75) -- +(1.5cm,-0.75cm);

\node at (1.9,0) {\small $\mu_2$};
\node at (1.9,1.5) {\small $\mu_1$};

\draw[->] (2.2,0) -- +(1cm,0);
\draw[->] (2.2,1.5) -- +(1cm,0);

\end{tikzpicture}
\vspace{0.3cm}
        \caption{Coded transmission.}
        \label{fig:code}
    \end{subfigure}	
     \caption{Depiction of the four transmission schemes.}
 \label{fig:schemes}
\end{figure}
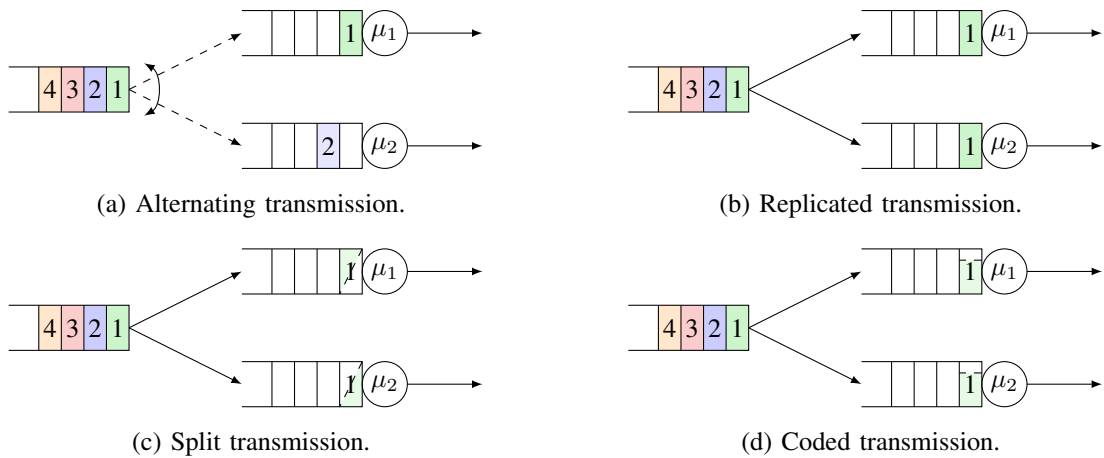

We consider a wireless \gls{vr} system, in which rendered frames of constant size are generated with a constant period $\tau$. The generated frames are then delivered to a user through a multipath connection using two different \glspl{rat}. We model the two paths as two separate queuing systems with Markovian service. The two paths, dubbed $1$ and $2$, then have exponential service times, with average rates $\mu_1$ and $\mu_2$, respectively: the service time is exponentially distributed to account for physical and medium access issues, which introduce a significant volatility in the delivery of the data. We also consider them to have infinite queues with \gls{fcfs} service. In the following, we will refer to the individual paths in the connection as ``paths'' or ``channels'' interchangeably, using ``system'' to indicate the overall multipath connection, which is then similar to a $D/M/2$ queue, as the arrival process is deterministic and the service process is Markovian for both servers, but the two paths have separate queues and the transmitter has a scheduler, i.e., it can choose which connection to send the data through. We assume that the two paths also function as \glspl{pec}, with erasure rates $\varepsilon_1$ and $\varepsilon_2$, respectively. We define the load $\rho_j$ on each path $j$ as:
\begin{equation}
 \rho_j=\frac{L\lambda_j}{\mu_j} \;\;\;j = 1,2,
\end{equation}
where $\lambda_j$ is the average arrival rate on the path, which depends on the scheduling process, and $L$ is the normalized size of the packets: a full frame is considered to have $L=1$. We consider four different scheduling strategies, which are depicted in Fig.~\ref{fig:schemes}: 
\begin{itemize}
 \item The \emph{alternating} transmission scheme, shown in Fig.~\ref{fig:alt}, is a simple round-robin scheduler: odd-numbered packets are scheduled on the first path, while even-numbered ones are scheduled on the second one. In this case, the multipath connection is used to reduce the load on the two paths: the arrival rate is $\lambda_j=\frac{1}{2\tau}$ for either connection, while we have $L=1$.
 \item The \emph{replicated} transmission scheme, shown in Fig.~\ref{fig:rep}, duplicates each frames in two packets, which are sent over both paths at once. Naturally, this increases the load on the system, as we have $\lambda_j=\frac{1}{\tau}$ on either path and $L=1$, but it also provides error protection, as an erasure on either path can be compensated by the other replica. 
 \item The \emph{split} transmission scheme, shown in Fig.~\ref{fig:split}, divides the frame in two smaller packets, each of which is sent over one path. In this case, the interarrival period is $\tau$ but the packet size is halved: we still have $\lambda_j=\frac{1}{\tau}$, but $L=0.5$.
 \item The \emph{coded} transmission scheme, shown in Fig.~\ref{fig:code}, is a hybrid between the replicated and split schemes: the transmitter sends two packets of equal size on each path. By using \gls{mdc}, each packet is decodable individually, and provides a lower-quality representation of the frame. If both packets are delivered, the video frame can be displayed at the full quality. The quality and size of each individual packet depend on the coding rate $\eta$: a value $\eta=0.5$ corresponds to the replicated scheme, while $\eta=1$ corresponds to the split scheme. The arrival rate at each path is still $\lambda_j=\frac{1}{\tau}$, while we have $L=\frac{1}{2\eta}$, as the frame is encoded and then split over the two paths.
\end{itemize}
In all schemes but the first one, arrivals on the two paths are synchronized, i.e., one packet is sent on each path for each frame. 
Naturally, more intelligent schedulers are possible: a more advanced version of the alternating scheme would schedule the frame based on the occupation of each queue, always sending packets on the less congested path. However, this kind of scheduler is difficult to study analytically, and we will only provide simulation results for it. We can also think of an adaptive coded scheme, which changes the \gls{mdc} coding rate depending on the state of the two queues, but this is beyond the scope of this work, where we aim to study the basic features of the four alternatives from Fig.~\ref{fig:schemes}.

In the following, we will represent random variables using capital letters, and their values with lower-case letters. Vectors are represented in bold, 
We define the state of the overall system when the $i$-th frame is generated as $\mathbf{Q}_i=(Q_{i,1},Q_{i,2})$, where $Q_{i,j}$ represents the number of packets in channel $j$, including the one in service. We derive next the system time and peak age distributions for the four transmission schemes, first for the alternating scheme, then for the coding-based ones.

\section{Analysis: Alternating transmission}\label{sec:an_alt}
In this scheme, packets are divided between the two paths in a round-robin fashion. In this case, the system is simple: each of the two paths is independent from the other and can be treated individually, with an arrival process with only half the rate. We can consider the error-free alternating $D/M/2$ system, in which the interarrival period for each path is $2\tau$. We then have the following distribution of the state $q_{i,j}$ just before an arrival, following the well-known formula derived by Erlang~\cite{erlang1917losning}:
\begin{equation}
 p_{Q_{i,j}}(q_{i,j})=(1-\sigma_j)\sigma_j^{q_{i,j}},\label{eq:p_md1_alt}
\end{equation} 
where the parameter $\sigma_j$ is the solution in $(0,1)$ to the following equation:
\begin{equation}
 x=e^{2\mu_j \tau(x-1)}.\label{eq:sigma}
\end{equation}
The existence of $\sigma_j$ is guaranteed if the paths are stable, i.e., if $\rho_j=\frac{1}{2\mu_j \tau}<1\,\forall j\in\{1,2\}$.
Since the two paths can be treated as independent, the \gls{pdf} of the system time $T_{i,j}$ for a successfully received packet on path $j$, is given by:
\begin{equation}
 p_{T_{i,j}}(t)=\mu_j(1-\sigma_j)e^{-\mu_j(1-\sigma_j)t}u(t).
\end{equation}
The \gls{pdf} of the system time $T_1^{\text{alt}}$ for successfully received packets is then given by:
\begin{equation}
 p_{T_1^{\text{alt}}}(t)=\sum_{q_{1,1}=0}^\infty (1-\sigma_1)\sigma_1^{q_{i,1}}\frac{\mu_1^{q_{i,1}+1}t^{q_{i,1}}e^{-\mu_1t}}{q_{i,1}!},
\end{equation}
and the same goes for $p_{T_1^{\text{alt}}}(t)$, with inverted indices. Solving the series, we get:
\begin{equation}
 p_{T_i^{\text{alt}}}(t)=\begin{cases}
            (1-\sigma_1)\mu_1e^{-\mu_1(1-\sigma_1)t}&\text{if $i$ is odd;}\\
            (1-\sigma_2)\mu_2e^{-\mu_2(1-\sigma_2)t} &\text{if $i$ is even.}
            \end{cases}
\end{equation}
If we take a random packet, we will simply get:
\begin{equation}
 p_{T^{\text{alt}}}(t)=\frac{p_{T_1^{\text{alt}}}(t)+p_{T_2^{\text{alt}}}(t)}{2}.
\end{equation}

For the \gls{aoi} calculation, it is convenient to work with the probability of a packet being relevant, i.e., a successive packet has not arrived before it. We calculate first the case without channel errors. If $i$ is odd, the probability $p_{r,1}$ that packet $i$ is relevant is given by:
\begin{equation}
 \begin{aligned}
  p_{r,1}=&P_{T_1}(\tau)+\int_\tau^\infty p_{T_1}(t)(1-P_{T_2}(t-\tau)) dt\\
  =&1-e^{-\mu_1(1-\sigma_1)\tau}\left(1-\frac{\mu_1(1-\sigma_1)}{\mu_1(1-\sigma_1)+\mu_2(1-\sigma_2)}\right).
 \end{aligned}
\end{equation}

The same is true for $p_{r,2}$, with inverted indices. The distribution of the peak age when a relevant packet arrives from path 1 is then:
\begin{equation}
  \begin{aligned}
   p_{\Delta,1}(\delta)=&\frac{(1-\sigma_1)}{p_{r,1}}u(\delta-\tau)\mu_1e^{-\mu_1(1-\sigma_1)(\delta-\tau)}\bigg[u(2\tau-\delta)(1-e^{-\mu_2(1-\sigma_2)\delta})\\
   &+u(\delta-2\tau)\bigg(e^{-\mu_2(1-\sigma_2)(\delta-\tau)}e^{\mu_1(1-\sigma_1)\tau}+(1-e^{-2\mu_2(1-\sigma_2)\tau})e^{-\mu_2(\delta-2\tau)}
   \\
   &+\frac{(1-\sigma_2)}{\sigma_2}e^{-\mu_2\delta}(e^{\mu_2\sigma_2\delta}-e^{2\mu_2\sigma_2\tau})\bigg)\bigg].
   \end{aligned}
\end{equation}

The unconditioned \gls{paoi} distribution is:
\begin{equation}
\begin{aligned}
 p_{\Delta^{\text{alt}}}(\delta)=&\frac{p_{r,1}p_{\Delta,1}(\delta)+p_{r,2}p_{\Delta,2}(\delta)}{p_{r,1}+p_{r,2}}.    
\end{aligned}
\end{equation}
If we consider error-prone channels, the exact \gls{paoi} distribution is hard to compute, as we need to consider all possible orderings of past packets. We then compute a lower bound, assuming that, for the transmission of packet $i$, packet $i-3$ has already arrived at the other path, i.e., multiple packets can be lost, but we never have more than 2 consecutive packets arrive out-of-order. This approximation is tight for low traffic scenarios. We denote the number of consecutive failures as $F$. If the number of consecutive failures is larger than 1, the bound is given by:
\begin{equation}
 p_{\Delta_1^{\text{alt},e}|F}(\delta|f)=p_{T_1|r_1}(\delta-(f+1)\tau)\,\forall f\geq 1,
\end{equation}

where the conditioned \gls{pdf} $p_{T_1|r_1}(t)$ is given by:
\begin{equation}
 p_{T_1|r_1}(t)=\frac{\mu_1(1-\sigma_1)e^{-\mu_1(1-\sigma_1)t}(\varepsilon_2+(1-\varepsilon_2)(1+u(t-\tau)(e^{-\mu_2(1-\sigma_2)(t-\tau)}-1))}{\varepsilon_2+(1-\varepsilon_2)p_{r,1}}.
\end{equation}
The same is true for the second path, after inverting the indices.
We can then compute $p_{\Delta^{\text{alt},e}}$ by using the law of total probability:
\begin{equation}
\begin{aligned}
 p_{\Delta^{\text{alt},e}}(\delta)=&\frac{(1-\varepsilon_1)\left((1-\varepsilon_2)p_{r,1}p_{\Delta_1}(\delta)+\sum_{f=1}^{\left\lfloor\frac{\delta}{\tau}\right\rfloor} p_{T_1|r_1}(\delta-(f+1)\tau)\varepsilon_1^{\left\lfloor{f}{2}\right\rfloor}\varepsilon_2^{\left\lceil{f}{2}\right\rceil}\right)}{(1-\varepsilon_1)(\varepsilon_2+(1-\varepsilon_2)p_{r,1})+(1-\varepsilon_2)(\varepsilon_1+(1-\varepsilon_1)p_{r,2})}\\
 &+\frac{(1-\varepsilon_2)\left((1-\varepsilon_1)p_{r,2}p_{\Delta_2}(\delta)+\sum_{f=1}^{\left\lfloor\frac{\delta}{\tau}\right\rfloor} p_{T_2|r_2}(\delta-(f+1)\tau)\varepsilon_1^{\left\lceil{f}{2}\right\rceil}\varepsilon_2^{\left\lfloor{f}{2}\right\rfloor}\right)}{(1-\varepsilon_1)(\varepsilon_2+(1-\varepsilon_2)p_{r,1})+(1-\varepsilon_2)(\varepsilon_1+(1-\varepsilon_1)p_{r,2})}.
 \end{aligned}
\end{equation}

\section{Analysis: Multipath transmission}\label{sec:an_mult}

In a synchronized $D/M/2$ system, frames arrive simultaneously at both paths with constant interarrival period $\tau$. This scenario can represent a multipath transmission, using a split, replicated, or coded approach, and the three approaches differ only by the coding rate $\eta$. The steady-state distribution for this kind of system was derived in~\cite{pinotsi2005synchronized} following Palm probability theory~\cite{baccelli2012palm}. The joint distribution of $\mathbf{Q}$, considering the overall system just before packet $i$ is generated, is given by:
\begin{equation}
 p_{\mathbf{Q}_i}(\mathbf{q}_i)=(1-\sigma_1)(1-\sigma_2)\sigma_1^{q_{i,1}}\sigma_2^{q_{i,2}},\label{eq:p_md1}
\end{equation} 
where $\sigma_j$ is defined, like in~\eqref{eq:sigma}, as the solution in $(0,1)$ of the following equation:
\begin{equation}
  x=e^{2\eta\mu_j \tau(x-1)}.\label{eq:sigma_eta}
\end{equation}
Naturally, the stability condition is now $\rho_j=\frac{1}{2\eta\mu_j\tau}<1\,\forall j\in\{1,2\}$. In the following, we compute the distribution of the system time for the first and last packet to arrive, which will be used for the system time calculation for the three multipath schemes.

\subsection{Minimum System Time Distribution (Error-Free)}

We now consider the $i$-th generated request, which causes two packets to arrive simultaneously at both paths at time $g_i$. The two packets will then arrive at the receiver at times $r_{i,1}$ and $r_{i,2}$, depending on the system time of the two paths. We consider the minimum system time $T_i^{\min}$, i.e., the system time of the first packet to be received:
\begin{equation}
 T_i^{\min}=\min_{j\in\{1,2\}}r_{i,j}-g_i.
\end{equation}
The minimum system time is then a random variable whose distribution is the minimum between the distributions of the two independent paths. This corresponds to the delay for a successful transmission in a replicated scheme, as the first copy of the frame to arrive is decoded and shown to the user.

We first solve the following series, which is used several times in the derivation of the system time \gls{pdf}. As the factor $\sigma$ is guaranteed to be in the open interval $(0,1)$ due to the stability condition on the two paths, its geometric series converges and the inversion of the two summations is possible:
\begin{equation}
 \begin{aligned}
  \sum_{q=1}^\infty \sigma^q\sum_{n=0}^q\frac{\alpha^{n}}{n!}&=\sum_{q=1}^\infty\sigma^q+\sum_{n=1}^\infty\frac{\alpha^n}{n!}\sum_{q=n-1}^\infty \sigma^q\\
  &=\frac{1}{1-\sigma}+\sum_{n=1}^\infty\frac{(\alpha\sigma)^n}{(1-\sigma)n!}\\
  &=\frac{e^{\alpha\sigma}}{1-\sigma}-1.\label{eq:series_sol}
 \end{aligned}
\end{equation}

We can now compute the conditioned \gls{pdf} of the system time distribution, considering the system state $\mathbf{q}_i$ as known. In the following, we omit the index of the packet $i$ for the sake of readability. We then distinguish four cases, depending on whether each queue is empty. In the first case, the two paths are empty, and the system time is the minimum between the service times at the two paths, which are two independent exponentially distributed random variables:
\begin{equation}
\begin{aligned}
 p_{T^{\min}|\mathbf{Q}}(t|(0,0))=&p_{T_{1}|Q_1}(t|0)(1-P_{T_{2}|Q_2}(t|0))+(1-P_{T_{1}|Q_1}(t|0))p_{T_{2}|Q_2}(t|0)\\
 =&2\eta(\mu_1+\mu_2)e^{-2\eta(\mu_1+\mu_2)t}u(t).\label{eq:t_both_empty_min}
 \end{aligned}
\end{equation}
In the second case, in which the first path is empty but the second has at least one packet in the queue, the system time is the minimum between an exponentially distributed random variable and an Erlang distributed one:
\begin{equation}
\begin{aligned}
 p_{T^{\min}|\mathbf{Q}}(t|(0,q_2))=&p_{T_{1}|Q_1}(t|0)(1-P_{T_{2}|Q_2}(t|q_2))+(1-P_{T_{1}|Q_1}(t|0))p_{T_{2}|Q_2}(t|q_2)\\
 =& \left[e^{-2\eta\mu_1 t}\frac{(2\eta\mu_2)^{q_{2}+1}t^{q_{2}}e^{-2\eta\mu_2 t}}{q_{2}!}+2\eta\mu_1 e^{-2\eta\mu_1 t}\sum_{n=0}^{q_{2}}\frac{(2\eta\mu_2 t)^n e^{-2\eta\mu_2 t}}{n!}\right]u(t).\label{eq:t_one_empty_min}
 \end{aligned}
\end{equation}
The third case, in which the second time is empty but the first one is not, is symmetrical to the second case, and the conditioned \gls{pdf} in this case is given by~\eqref{eq:t_one_empty_min}, inverting the indices of the two paths. Finally, if both paths have packets in the queue, the resulting distribution is the minimum between two Erlang distributed variables:
\begin{equation}
\begin{aligned}
 p_{T^{\min}|\mathbf{Q}}(t|(q_{1},q_{2}))=&p_{T_{1}|Q_1}(t|q_1)(1-P_{T_{2}|Q_2}(t|q_2))+(1-P_{T_{1}|Q_1}(t|q_1))p_{T_{2}|Q_2}(t|q_2)\\
 =&\Bigg[\frac{(2\eta\mu_2)^{q_{2}+1}t^{q_{2}}e^{-2\eta\mu_2 t}}{q_{2}!}\sum_{n=0}^{q_{1}}\frac{(2\eta\mu_1 t)^n e^{-2\eta\mu_1 t}}{n!}\\
 &+\frac{(2\eta\mu_1)^{q_{1}+1}t^{q_{1}}e^{-2\eta\mu_1 t}}{q_{1}!}\sum_{n=0}^{q_{2}}\frac{(2\eta\mu_2 t)^n e^{-2\eta\mu_2 t}}{n!}\Bigg]u(t).
 \end{aligned}\label{eq:t_full_min}
\end{equation}
We now apply the law of total probability to remove the condition on $Q_{2}$, using the solution of the series in~\eqref{eq:series_sol}:
\begin{equation}
\begin{aligned}
  p_{T^{\min}|Q_1}(t|q_{1})=&\sum_{q_{2}=0}^\infty \frac{p_{\mathbf{Q}}((q_{1},q_2))}{\sum_{k=0}^\infty p_{\mathbf{Q}}((q_{1},k))}p_{T^{\min}|\mathbf{Q}}(t|(q_{1},q_{2}))\\
  =&u(t)(1-\sigma_2)\sum_{\mathclap{q_{2}=0}}^\infty\sigma_2^{q_{2}}\Bigg[\frac{(2\eta\mu_2)^{q_2+1}t^{q_2}e^{-2\eta\mu_2 t}}{q_{2}!}\sum_{n=0}^{q_{1}}\frac{(2\eta\mu_1 t)^n e^{-2\eta\mu_1 t}}{n!}\\
  &+\frac{(2\eta\mu_1)^{q_1+1}t^{q_{1}}e^{-2\eta\mu_1 t}}{q_{1}!}\sum_{n=0}^{q_{2}}\frac{(2\eta\mu_2 t)^n e^{-2\eta\mu_2 t}}{n!}\Bigg]\\
=&u(t)(1-\sigma_2)e^{-2\eta(\mu_1+\mu_2)t}\left[2\eta\mu_2 e^{2\eta\mu_2\sigma_2 t}\sum_{n=0}^{q_1}\frac{(2\eta\mu_1 t)^n}{n!}+\frac{(2\eta\mu_1)^{q_1+1}t^{q_1}}{q_1!}\frac{e^{2\eta\mu_2\sigma_2 t}}{1-\sigma_2}\right].\\
\end{aligned} \label{eq:t_one_full_uncond_min}
\end{equation}

Finally, we remove the condition on $Q_1$ using the same principles and get the unconditioned \gls{pdf} of the minimum system time:
\begin{equation}
\begin{aligned}
  p_{T^{\min}}(t)=&\sum_{q_{1}=0}^\infty p_{T^{\min}|Q_1}(t|q_{1}) p_{\mathbf{Q}_{1}}(q_{1})\\
  =&u(t)e^{-2\eta(\mu_1+\mu_2)t}(1-\sigma_1)(1-\sigma_2)\sum_{\mathclap{q_1=0}}^\infty\sigma_1^{q_1}e^{2\eta\mu_2\sigma_2 t}\Bigg[\!2\eta\mu_2\sum_{\mathclap{n=0}}^{q_1}\frac{(2\eta\mu_1 t)^n}{n!}+\frac{(2\eta\mu_1)^{q_1+1}t^{q_1}}{(1-\sigma_2)q_1!}\!\Bigg]\\
   =&u(t)2\eta\left[\mu_1(1-\sigma_1)+\mu_2(1-\sigma_2)\right]e^{-2\eta(\mu_1(1-\sigma_1)+\mu_2(1-\sigma_2))t}.
\end{aligned}
\end{equation}
As stated above, the minimum system time $T^{\min}$ can represent the latency of a replicated packet sent over multiple connections, only one copy of which needs to be delivered in order to fulfill the request. It is interesting to note that the resulting system time is still exponentially distributed, with parameter $2\eta(\mu_1(1-\sigma_1)+\mu_2(1-\sigma_2))$. The \gls{cdf} of the system time distribution is then given by:
\begin{equation}
 P_{T^{\min}}(t)=u(t)(1-e^{-2\eta(\mu_1(1-\sigma_1)+\mu_2(1-\sigma_2))t}).
\end{equation}

\subsection{Minimum System Time Distribution (Error-Prone)}
We can now consider a more interesting scenario, in which the two channels are error-prone and behave as independent \glspl{pec} with packet erasure probability $\varepsilon_1$ and $\varepsilon_2$, respectively. This is a key scenario for replicated transmission, as sending two copies of a frame can protect it from channel errors as well as reducing latency.
We then have a probability $p_s^{\min}=1-\varepsilon_1\varepsilon_2$ of fulfilling each request. Furthermore, we define $\phi_{1,2}=\frac{(1-\varepsilon_1)(1-\varepsilon_2)}{p_s^{\min}}$, $\phi_1=\frac{(1-\varepsilon_1)\varepsilon_2}{p_s^{\min}}$, and $\phi_2=\frac{(1-\varepsilon_2)\varepsilon_1}{p_s^{\min}}$, to identify the probability of having a successful request with two received packet, only the packet on the first path being received, and only the packet on the second path being received, respectively. We now adjust the system time \gls{pdf} calculation to account for the possibility of erasure. 
We denote the minimum system time with errors as $T^{\min,e}$, and divide the computation in the same four cases as above. We now consider the conditioned system time \gls{pdf} when both paths are empty:
\begin{equation}
 p_{T^{\min,e}|\mathbf{Q}}(t|(0,0))=\phi_{1,2}p_{T^{\min}|\mathbf{Q}}(t|(0,0))+2\eta(\phi_1\mu_1e^{-2\eta\mu_1t}+\phi_2\mu_2e^{-2\eta\mu_2t})u(t).\label{eq:t_both_empty_min_err}
\end{equation}
In the second case, in which the first path is empty but the second has at least one packet in the queue, the conditioned system time \gls{pdf} is given by:
\begin{equation}
\begin{aligned}
  p_{T^{\min,e}|\mathbf{Q}}(t|(0,q_{2}))=&\phi_{1,2}p_{T^{\min}|\mathbf{Q}}(t|(0,q_{2}))+2\eta\phi_1\mu_1 e^{-2\eta\mu_1 t}u(t)+\phi_2\frac{(2\eta\mu_2)^{q_{2}+1}t^{q_{2}}e^{-2\eta\mu_2 t}}{q_{2}!}u(t).
  \label{eq:t_one_empty_min_err}
\end{aligned}
\end{equation}
As for the error-free system, the third case, in which the second time is empty but the first one is not, is symmetrical to the second case. Finally, if both paths have packets in the queue, the resulting distribution is:
\begin{equation}
\begin{aligned}
 p_{T^{\min,e}|\mathbf{Q}}(t|(q_{1},q_{2}))=&\phi_{1,2}p_{T^{\min}|\mathbf{Q}}(t|(q_{1},q_{2}))\\
 &+\phi_1\frac{(2\eta\mu_1)^{q_{1}+1}t^{q_{1}}e^{-2\eta\mu_1 t}}{q_{1}!}u(t)+\phi_2\frac{(2\eta\mu_2)^{q_{2}+1}t^{q_{2}}e^{-2\eta\mu_2 t}}{q_{2}!}u(t).
 \end{aligned}\label{eq:t_full_min_err}
\end{equation}
As above, we now use the law of total probability to remove the condition on $Q_{2}$:
\begin{equation}
\begin{aligned}
  p_{T^{\min,e}|Q_1}(t|q_{1})=&\sum_{q_{2}=0}^\infty \frac{p_{\mathbf{Q}}((q_{1},q_2))}{\sum_{k=0}^\infty p_{\mathbf{Q}}((q_{1},k))}p_{T^{\min,e}|\mathbf{Q}}(t|(q_{1},q_{2}))\\
=&(1-\sigma_2)u(t)\Bigg[\phi_{1,2}p_{T^{\min}|Q_1}(t|q_{1})+\phi_1\frac{(2\eta\mu_1)^{q_{1}+1}t^{q_{1}}e^{-2\eta\mu_1 t}}{q_{1}!}
\\&+2\eta\phi_2\mu_2e^{-2\eta\mu_2t}(e^{2\eta\mu_2\sigma_2t}-1)\Bigg].
\end{aligned} \label{eq:t_one_full_uncond_min_err}
\end{equation}
Finally, we remove the condition on $Q_1$ using the same principles and get the unconditioned \gls{pdf} of the minimum system time:
\begin{equation}
\begin{aligned}
  p_{T^{\min,e}}(t)=&\sum_{q_{1}=0}^\infty p_{T^{\min,e}|Q_1}(t|q_{1}) p_{\mathbf{Q}_{1}}(q_{1})\\
   =&\phi_{1,2}p_{T^{\min}}(t)+2\eta(1-\sigma_1)(1-\sigma_2)u(t)\left[\phi_1\mu_1e^{-2\eta\mu_1(1-\sigma_1)t}+\phi_2\mu_2e^{-2\eta\mu_2(1-\sigma_2)t}\right].
\end{aligned}
\end{equation}
We remind the reader that the latency above is computed only for successful requests, while a fraction $1-p_s^{\min}$ of the frames is lost, as both copies of the packet are erased.
The system time \gls{cdf} in the error-prone case is given by:
\begin{equation}
 P_{T^{\min,e}}(t)=\phi_{1,2}P_{T^{\min}}(t)+u(t)2\eta\big[\phi_1(1-\sigma_2)(1-e^{-2\eta\mu_1(1-\sigma_1)t})+\phi_2(1-\sigma_1)(1-e^{-2\eta\mu_2(1-\sigma_2)t})\big].
\end{equation}

\subsection{Maximum System Time Distribution}

We now consider the maximum system time $T_i^{\max}$, i.e., the system time of the second packet to be received:
\begin{equation}
 T_i^{\max}=\max_{j\in\{1,2\}}r_{i,j}-g_i.
\end{equation}
In a split transmission scheme, the maximum system time corresponds to the delay of a sucessful frame. As for $T_i^{\min}$, the distribution of $T_i^{\max}$ is the maximum of the two distributions of the system times at the two paths, which are independent if the state of the two queues is given. The distribution we compute below is only for successful requests, which in this case require both packets to be delivered: we have $p_s^{\max}=(1-\varepsilon_1)(1-\varepsilon_2)$. The value of $\sigma_j$ is given by~\eqref{eq:sigma_eta}. As above, we omit the index of the request $i$ for brevity, and consider four different cases based on the state of the two queues immediately before the request. In the first case, both queues are empty, and the delay is the maximum between two exponential random variables:
\begin{equation}
\begin{aligned}
 p_{T^{\max}|\mathbf{Q}}(t|(0,0))=2\eta\left(\mu_1 e^{-2\eta\mu_1 t}+\mu_2 e^{-2\eta\mu_2 t}-(\mu_1+\mu_2)e^{-2\eta(\mu_1+\mu_2)t}\right)u(t).
 \end{aligned}\label{eq:t_both_empty}
\end{equation}
In the second case, the second path has at least one packet in the queue when the request arrives, while the first one is empty. In this case, the system time distribution is the maximum between an exponentially distributed random variable and an Erlang distributed one:
\begin{equation}
\begin{aligned}
 p_{T^{\max}|\mathbf{Q}}(t|(0,q_{2}))=&\Bigg[(1-e^{-2\eta\mu_1 t})\frac{(2\eta\mu_2)^{q_{2}+1}t^{q_{2}}e^{-2\eta\mu_2 t}}{q_{2}!}
 \\&+2\eta\mu_1 e^{-2\eta\mu_1 t}\left(1-\sum_{n=0}^{q_{2}}\frac{(2\eta\mu_2 t)^n e^{-2\eta\mu_2 t}}{n!}\right)\Bigg]u(t).\label{eq:t_one_empty}
 \end{aligned}
\end{equation}
As for the minimum system time, $p_{T^{\max}|\mathbf{Q}}(t|(q_{1},0))$ follows the same distribution, inverting the indices of the two paths.
Finally, we consider the case in which both paths have queued packets. In this case, the distribution of the overall system time is the maximum between two Erlang distributed variables, which correspond to the system time for the two paths:
\begin{equation}
\begin{aligned}
 p_{T^{\max}|\mathbf{Q}}(t|(q_{1},q_{2}))=&\left(1-\sum_{n=0}^{q_{1}}\frac{(2\eta\mu_1 t)^n e^{-2\eta\mu_1 t}}{n!}\right)\frac{(2\eta\mu_2)^{q_{2}+1}t^{q_{2}}e^{-2\eta\mu_2 t}}{q_{2}!}u(t)\\
 &+\frac{(2\eta\mu_1)^{q_{1}+1}t^{q_{1}}e^{-2\eta\mu_1 t}}{q_{1}!}\left(1-\sum_{n=0}^{q_{2}}\frac{(2\eta\mu_2 t)^n e^{-2\eta\mu_2 t}}{n!}\right)u(t).
 \end{aligned}\label{eq:t_full}
\end{equation}
We can then use~\eqref{eq:t_both_empty},~\eqref{eq:t_one_empty}, and~\eqref{eq:t_full} to remove the condition on $Q_2$ from the system time distribution, using the law of total probability to condition only on the first queue's state:
\begin{equation}
\begin{aligned}
  p_{T^{\max}|Q_1}(t|q_{1})=&\sum_{q_{2}=0}^\infty \frac{p_{\mathbf{Q}}((q_{1},q_2))}{\sum_{k=0}^\infty p_{\mathbf{Q}}((q_{1},k))}p_{T^{\max}|\mathbf{Q}}(t|(q_{1},q_{2}))\\
=&u(t)(1-\sigma_2)\Bigg[2\eta\mu_2 e^{-2\eta\mu_2 (1-\sigma_2)t}\left(1-\sum_{n=0}^{q_{1}}\frac{(2\eta\mu_1 t)^n e^{-2\eta\mu_1 t}}{n!}\right)\\
&+\frac{(2\eta\mu_1)^{q_{1}+1}t^{q_{1}}e^{-2\eta\mu_1 t}}{(1-\sigma_2)q_{1}!}(1-e^{-2\eta\mu_2(1-\sigma_2)t}\Bigg].
\end{aligned} \label{eq:t_one_full_uncond}
\end{equation}
We can now remove the condition on the state of the first queue as well:
\begin{equation}
\begin{aligned}
  p_{T^{\max}}(t)=&\sum_{q_{1}=0}^\infty p_{T^{\max}|Q_1}(t|q_{1}) p_{\mathbf{Q}_{1}}(q_{1})\\
  =&u(t)\!2\eta\big[(1-\sigma_1)\mu_1 e^{-2\eta\mu_1(1-\sigma_1)t}(1-e^{-2\eta\mu_2(1-\sigma_2)t})\\
  &+(1-\sigma_2)\mu_2 e^{-2\eta\mu_2(1-\sigma_2)t}(1-e^{-2\eta\mu_1(1-\sigma_1)t})\big].
\end{aligned}
\end{equation}
The maximum system time represents the latency for a system which requires both packets to fulfill the request. Network coded transmissions, in which both packets contain part of the information required at the receiver, is one example of this kind of system. The \gls{cdf} of the system time is:
\begin{equation}
\begin{aligned}
  P_{T^{\max}}(t)=u(t)\left[1-e^{-2\eta\mu_1(1-\sigma_1)t}-e^{-2\eta\mu_2(1-\sigma_2)t}+e^{-2\eta(\mu_1(1-\sigma_1)+\mu_2(1-\sigma_2))t}\right].
\end{aligned}
\end{equation}

\subsection{System time and PAoI for synchronized systems}

After computing $p_T^{\min,e}$ and $p_T^{\max}$, we can go ahead and get the system time for a coded system. In the most general case, we have a coded system with $\eta\in(0.5,1)$. We then have the following \gls{cdf}:
\begin{align}
 p_{T^{\text{LQ}}}(t)=p_T^{\min,e}(t)\\
 p_{T^{\text{HQ}}}(t)=p_T^{\max,e}(t),
\end{align}
where $T^{\text{LQ}}$ is the delay of the low-quality version of the frame and $T^{\text{HQ}}$ is the delay of the high-quality version of the frame. The split and replicated systems are two extreme cases of the coded transmission: in the former, the low-quality frame is undecodable, and $\eta=1$, while in the latter, the low quality version is identical to the full-quality one, and $\eta=0.5$.

We can now consider the distribution of the \gls{paoi}, denoted by $\Delta$, in the systems for which we just computed the system time \gls{pdf}. For simplicity, we refer to the more general error-prone system, which the error-free case is a special case of with $\varepsilon_1=\varepsilon_2=0$. If we have a synchronized $D/M/2$ system with success probability $p_s$ and system time \gls{pdf} $p_T(t)$ for successful packets, we need to consider the possibility of request failures. We denote the number of failures as $F$: if there are $f$ consecutive failures, the \gls{pdf} of the \gls{paoi} is simply given by:
\begin{equation}
 p_{\Delta|F}(\delta|f)=p_{T}(\delta-(f+1)\tau)u(\delta-(f+1)\tau) \;\;\;\;\forall f \geq 1.
\end{equation}
This is due to the deterministic nature of the arrival process, which increases the age by $\tau$ for every failure. As failures are independently distributed and the success probability is $p_s$, we can now apply the law of total probability to remove the condition:
\begin{equation}
\begin{aligned}
 p_{\Delta}(\delta)=p_s\sum_{f=0}^{\left\lfloor\frac{\delta}{\tau}\right\rfloor-1}(1-p_s)^f p_T(\delta-(f+1)\tau).\label{eq:paoi}
\end{aligned}
\end{equation}
By substituting $p_s^{\min}$ and $p_T^{\min,e}$, or $p_s^{\max}$ and $p_T^{\max}$, we can compute the distribution of the \gls{paoi} for replicated and split systems, respectively.

In a coded transmission scheme with \gls{mdc}, the concept of \gls{paoi} is split between the two qualities: as above, we can substitute $p_s^{\min}$ and $p_T^{\min,e}$ into~\eqref{eq:paoi} to compute the \gls{paoi} of low-quality frames, or $p_s^{\max}$ and $p_T^{\max}$ to compute the peak age considering only high-quality frames.

\section{Simulation Results}\label{sec:sim}

In this section, we compare our analytical results with an extensive Monte Carlo simulation, consisting of over $10^6$ packets, in order to verify the analysis.
The first 1000 packets of each simulation were cut from the results to avoid the initial transition effects and ensure that the system was only considered in a steady state. We analyzed both the system time for the four schemes and the \gls{paoi}, considering both the full distribution and the 99th percentile, used as a proxy for the worst-case performance.

\subsection{System time}

\begin{figure}
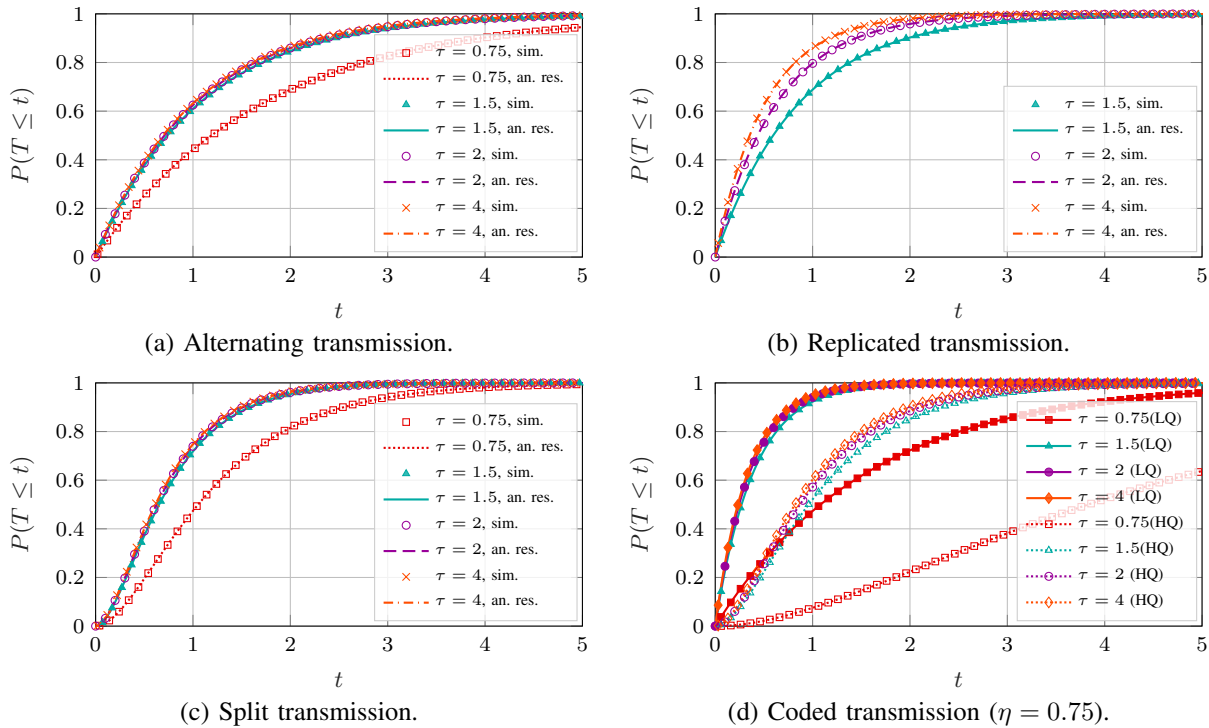

    \centering
	\begin{subfigure}[b]{.49\linewidth}
	    \centering
        \input{./figures/dm2_lat_dist_rho_alt.tex}\vspace{-0.7cm}
        \caption{Alternating transmission.}
        \label{fig:lat_dist_noerr_alt}
    \end{subfigure}	
	\centering
	\begin{subfigure}[b]{.49\linewidth}
	    \centering
        \input{./figures/dm2_lat_dist_rho_min.tex}\vspace{-0.7cm}
        \caption{Replicated transmission.}
        \label{fig:lat_dist_noerr_min}
    \end{subfigure}	
	\begin{subfigure}[b]{.49\linewidth}
	    \centering
        \input{./figures/dm2_lat_dist_rho_max.tex}\vspace{-0.7cm}
        \caption{Split transmission.}
        \label{fig:lat_dist_noerr_max}
    \end{subfigure}	
    	\begin{subfigure}[b]{.49\linewidth}
	    \centering
        \input{./figures/dm2_lat_dist_rho_code.tex}\vspace{-0.7cm}
        \caption{Coded transmission ($\eta=0.75$).}
        \label{fig:lat_dist_noerr_cod}
    \end{subfigure}	
     \caption{Latency \gls{cdf} for the four schemes in an error-free $D/M/2$ with $\mu_1=\mu_2=1$.}
 \label{fig:lat_CDF_bal}
\end{figure}

We first look at the latency of the four schemes in an error-free system. The two paths have the same service rates, which are set to 1. Fig.~\ref{fig:lat_CDF_bal} shows the \gls{cdf} of the system time for different values of the inter-frame period $\tau$.
The alternating transmission scheme in Fig.~\ref{fig:lat_dist_noerr_alt} can support $\tau=0.75$, but it has a long tail: this is due to the fact that it relies on a single system sending a large packet. The replicated scheme can avoid this problem, as shown in Fig.~\ref{fig:lat_dist_noerr_min}, as its latency is the minimum between the two paths. However, the higher load it imposes on the system means that it cannot support a frame rate higher than 1 (e.g., the $\tau = 0.75$ case represented in the other schemes). This is not true for the split scheme in Fig.~\ref{fig:lat_dist_noerr_max}, which has both a lower load and a light tail. While it requires both paths to deliver their packets to decode a frame, it can still have a good worst-case performance because of the reduced packet size. Finally, the coded scheme, whose performance is shown in Fig.~\ref{fig:lat_dist_noerr_cod}, is a compromise between replicating and splitting: it can deliver a lower-quality version of the frame much faster than the other systems, but the full-quality version has a high latency, since it requires both packets to be delivered and each packet is larger than in the split system. In this case, the high-quality curve for $\tau = 0.75$ is still in the stability region, unlike the replicated scheme, although with a long tail. As expected, the simulated results perfectly match the analytical curves. This is also true for the coded transmission, but Fig.~\ref{fig:lat_dist_noerr_cod} does not show the Monte Carlo results to improve clarity.

\begin{figure}
    \centering
	\begin{subfigure}[b]{.49\linewidth}
	    \centering
        \begin{tikzpicture}
\pgfplotsset{every tick label/.append style={font=\scriptsize}}
\tikzstyle{dotted}= [dash pattern=on \pgflinewidth off 0.5mm] 
\tikzstyle{dashed}= [dash pattern=on 7.5*0.8*0.8pt off 7.5*0.4*0.8pt]
\tikzstyle{dashdotted} = [dash pattern=on 7.5*0.8*0.6pt off 7.5*0.8*0.3pt on \the\pgflinewidth off 7.5*0.8*0.3pt]
\tikzstyle{dotted2} = [dash pattern=on 7.5*0.8*0.3pt off 7.5*0.8*0.2pt]

\begin{axis}[%
width=\sfwidth,
height=\sfheight,
at={(0,0)},
scale only axis,
xmin=0.5,
xmax=1,
xlabel near ticks,
xlabel style={font=\footnotesize\color{white!15!black}},
xlabel={$\eta$},
ylabel near ticks,
ymin=0,
ymax=10,
ylabel style={font=\footnotesize\color{white!15!black}},
ylabel={$T  _{99}$},
axis background/.style={fill=white},
xmajorgrids,
ymajorgrids,
legend style={font=\tiny, at={(0.01,0.02)}, anchor=south west, legend cell align=left, align=left, fill opacity=0.8, draw opacity=1, text opacity=1, draw=white!80!black}
]

\addplot [color=cyan, only marks,  mark=triangle*, mark size=1.5]
  table[row sep=newline]{
1 7.9067
};
\addlegendentry{Alternating}

\addplot [color=orange_D, only marks,  mark=x, mark size=2]
  table[row sep=newline]{
1 4.5433
};
\addlegendentry{Split}

\addplot [color=red, densely dotted, line width=0.8pt]
  table[row sep=newline]{
0.500000000000000	11
0.510000000000000	11
0.520000000000000	11
0.530000000000000	11
0.540000000000000	11
0.550000000000000	11
0.560000000000000	11
0.570000000000000	11
0.580000000000000	11
0.590000000000000	11
0.600000000000000	11
0.610000000000000	11
0.620000000000000	11
0.630000000000000	11
0.640000000000000	11
0.650000000000000	11
0.660000000000000	11
0.670000000000000	11
0.680000000000000	11
0.690000000000000	11
0.700000000000000	11
0.710000000000000	11
0.720000000000000	11
0.730000000000000	9.37320000000000
0.740000000000000	8.13250000000000
0.750000000000000	7.18920000000000
0.760000000000000	6.44780000000000
0.770000000000000	5.84970000000000
0.780000000000000	5.35690000000000
0.790000000000000	4.94390000000000
0.800000000000000	4.59260000000000
0.810000000000000	4.29020000000000
0.820000000000000	4.02710000000000
0.830000000000000	3.79600000000000
0.840000000000000	3.59150000000000
0.850000000000000	3.40920000000000
0.860000000000000	3.24570000000000
0.870000000000000	3.09810000000000
0.880000000000000	2.96430000000000
0.890000000000000	2.84240000000000
0.900000000000000	2.73080000000000
0.910000000000000	2.62840000000000
0.920000000000000	2.53390000000000
0.930000000000000	2.44660000000000
0.940000000000000	2.36560000000000
0.950000000000000	2.29020000000000
0.960000000000000	2.21990000000000
0.970000000000000	2.15420000000000
0.980000000000000	2.09260000000000
0.990000000000000	2.03480000000000
1	1.98040000000000
  };
\addlegendentry{Coded (LQ)}

\addplot [color=black, densely dashed, line width=0.8pt]
  table[row sep=newline]{
0.500000000000000	11
0.510000000000000	11
0.520000000000000	11
0.530000000000000	11
0.540000000000000	11
0.550000000000000	11
0.560000000000000	11
0.570000000000000	11
0.580000000000000	11
0.590000000000000	11
0.600000000000000	11
0.610000000000000	11
0.620000000000000	11
0.630000000000000	11
0.640000000000000	11
0.650000000000000	11
0.660000000000000	11
0.670000000000000	11
0.680000000000000	11
0.690000000000000	11
0.700000000000000	11
0.710000000000000	11
0.720000000000000	11
0.730000000000000	11
0.740000000000000	11
0.750000000000000	11
0.760000000000000	11
0.770000000000000	11
0.780000000000000	11
0.790000000000000	11
0.800000000000000	11
0.810000000000000	9.85560000000000
0.820000000000000	9.25050000000000
0.830000000000000	8.71910000000000
0.840000000000000	8.24880000000000
0.850000000000000	7.82950000000000
0.860000000000000	7.45340000000000
0.870000000000000	7.11400000000000
0.880000000000000	6.80620000000000
0.890000000000000	6.52580000000000
0.900000000000000	6.26920000000000
0.910000000000000	6.03360000000000
0.920000000000000	5.81640000000000
0.930000000000000	5.61550000000000
0.940000000000000	5.42910000000000
0.950000000000000	5.25580000000000
0.960000000000000	5.09410000000000
0.970000000000000	4.94300000000000
0.980000000000000	4.80140000000000
0.990000000000000	4.66840000000000
1	4.54330000000000
};
\addlegendentry{Coded (HQ)}

\end{axis}
\end{tikzpicture}
        \caption{Inter-frame period $\tau=0.75$.}
        \label{fig:lat_eta_D075}
    \end{subfigure}	
	\centering
	\begin{subfigure}[b]{.49\linewidth}
	    \centering
        \begin{tikzpicture}
\pgfplotsset{every tick label/.append style={font=\scriptsize}}
\tikzstyle{dotted}= [dash pattern=on \pgflinewidth off 0.5mm] 
\tikzstyle{dashed}= [dash pattern=on 7.5*0.8*0.8pt off 7.5*0.4*0.8pt]
\tikzstyle{dashdotted} = [dash pattern=on 7.5*0.8*0.6pt off 7.5*0.8*0.3pt on \the\pgflinewidth off 7.5*0.8*0.3pt]
\tikzstyle{dotted2} = [dash pattern=on 7.5*0.8*0.3pt off 7.5*0.8*0.2pt]

\begin{axis}[%
width=\sfwidth,
height=\sfheight,
at={(0,0)},
scale only axis,
xmin=0.5,
xmax=1,
xlabel near ticks,
xlabel style={font=\footnotesize\color{white!15!black}},
xlabel={$\eta$},
ylabel near ticks,
ymin=0,
ymax=10,
ylabel style={font=\footnotesize\color{white!15!black}},
ylabel={$T  _{99}$},
axis background/.style={fill=white},
xmajorgrids,
ymajorgrids,
legend style={font=\tiny, at={(0.99,0.98)}, anchor=north east, legend cell align=left, align=left, fill opacity=0.8, draw opacity=1, text opacity=1, draw=white!80!black}
]

\addplot [color=cyan, only marks,  mark=triangle*, mark size=1.5]
  table[row sep=newline]{
1 4.9016    
};
\addlegendentry{Alternating}

\addplot [color=violet, only marks,  mark=o, mark size=1.5]
  table[row sep=newline]{
0.5 3.9558
};
\addlegendentry{Replicated}

\addplot [color=orange_D, only marks,  mark=x, mark size=2]
  table[row sep=newline]{
1 2.8155
};
\addlegendentry{Split}

\addplot [color=red, densely dotted, line width=0.8pt]
  table[row sep=newline]{
0.500000000000000	3.95580000000000
0.510000000000000	3.75650000000000
0.520000000000000	3.57810000000000
0.530000000000000	3.41750000000000
0.540000000000000	3.27220000000000
0.550000000000000	3.14010000000000
0.560000000000000	3.01930000000000
0.570000000000000	2.90850000000000
0.580000000000000	2.80650000000000
0.590000000000000	2.71220000000000
0.600000000000000	2.62480000000000
0.610000000000000	2.54350000000000
0.620000000000000	2.46780000000000
0.630000000000000	2.39690000000000
0.640000000000000	2.33050000000000
0.650000000000000	2.26820000000000
0.660000000000000	2.20960000000000
0.670000000000000	2.15430000000000
0.680000000000000	2.10200000000000
0.690000000000000	2.05260000000000
0.700000000000000	2.00570000000000
0.710000000000000	1.96130000000000
0.720000000000000	1.91900000000000
0.730000000000000	1.87870000000000
0.740000000000000	1.84040000000000
0.750000000000000	1.80370000000000
0.760000000000000	1.76870000000000
0.770000000000000	1.73530000000000
0.780000000000000	1.70320000000000
0.790000000000000	1.67250000000000
0.800000000000000	1.64300000000000
0.810000000000000	1.61470000000000
0.820000000000000	1.58750000000000
0.830000000000000	1.56130000000000
0.840000000000000	1.53600000000000
0.850000000000000	1.51170000000000
0.860000000000000	1.48830000000000
0.870000000000000	1.46560000000000
0.880000000000000	1.44380000000000
0.890000000000000	1.42260000000000
0.900000000000000	1.40220000000000
0.910000000000000	1.38240000000000
0.920000000000000	1.36320000000000
0.930000000000000	1.34470000000000
0.940000000000000	1.32670000000000
0.950000000000000	1.30920000000000
0.960000000000000	1.29230000000000
0.970000000000000	1.27580000000000
0.980000000000000	1.25980000000000
0.990000000000000	1.24430000000000
1	1.22920000000000
  };
\addlegendentry{Coded (LQ)}

\addplot [color=black, densely dashed, line width=0.8pt]
  table[row sep=newline]{
0.500000000000000	9.08670000000000
0.510000000000000	8.62810000000000
0.520000000000000	8.21790000000000
0.530000000000000	7.84860000000000
0.540000000000000	7.51440000000000
0.550000000000000	7.21050000000000
0.560000000000000	6.93270000000000
0.570000000000000	6.67790000000000
0.580000000000000	6.44330000000000
0.590000000000000	6.22640000000000
0.600000000000000	6.02540000000000
0.610000000000000	5.83840000000000
0.620000000000000	5.66420000000000
0.630000000000000	5.50120000000000
0.640000000000000	5.34860000000000
0.650000000000000	5.20520000000000
0.660000000000000	5.07030000000000
0.670000000000000	4.94310000000000
0.680000000000000	4.82290000000000
0.690000000000000	4.70920000000000
0.700000000000000	4.60150000000000
0.710000000000000	4.49920000000000
0.720000000000000	4.40200000000000
0.730000000000000	4.30940000000000
0.740000000000000	4.22110000000000
0.750000000000000	4.13690000000000
0.760000000000000	4.05640000000000
0.770000000000000	3.97950000000000
0.780000000000000	3.90570000000000
0.790000000000000	3.83510000000000
0.800000000000000	3.76720000000000
0.810000000000000	3.70210000000000
0.820000000000000	3.63950000000000
0.830000000000000	3.57920000000000
0.840000000000000	3.52120000000000
0.850000000000000	3.46530000000000
0.860000000000000	3.41130000000000
0.870000000000000	3.35930000000000
0.880000000000000	3.30900000000000
0.890000000000000	3.26040000000000
0.900000000000000	3.21330000000000
0.910000000000000	3.16780000000000
0.920000000000000	3.12380000000000
0.930000000000000	3.08110000000000
0.940000000000000	3.03970000000000
0.950000000000000	2.99950000000000
0.960000000000000	2.96060000000000
0.970000000000000	2.92270000000000
0.980000000000000	2.88600000000000
0.990000000000000	2.85020000000000
1	2.81550000000000
};
\addlegendentry{Coded (HQ)}

\end{axis}
\end{tikzpicture}
        \caption{Inter-frame period $\tau=1.5$.}
        \label{fig:lat_eta_D15}
    \end{subfigure}	
	\begin{subfigure}[b]{.49\linewidth}
	    \centering
        \begin{tikzpicture}
\pgfplotsset{every tick label/.append style={font=\scriptsize}}
\tikzstyle{dotted}= [dash pattern=on \pgflinewidth off 0.5mm] 
\tikzstyle{dashed}= [dash pattern=on 7.5*0.8*0.8pt off 7.5*0.4*0.8pt]
\tikzstyle{dashdotted} = [dash pattern=on 7.5*0.8*0.6pt off 7.5*0.8*0.3pt on \the\pgflinewidth off 7.5*0.8*0.3pt]
\tikzstyle{dotted2} = [dash pattern=on 7.5*0.8*0.3pt off 7.5*0.8*0.2pt]

\begin{axis}[%
width=\sfwidth,
height=\sfheight,
at={(0,0)},
scale only axis,
xmin=0.5,
xmax=1,
xlabel near ticks,
xlabel style={font=\footnotesize\color{white!15!black}},
xlabel={$\eta$},
ylabel near ticks,
ymin=0,
ymax=10,
ylabel style={font=\footnotesize\color{white!15!black}},
ylabel={$T  _{99}$},
axis background/.style={fill=white},
xmajorgrids,
ymajorgrids,
legend style={font=\tiny, at={(0.99,0.98)}, anchor=north east, legend cell align=left, align=left, fill opacity=0.8, draw opacity=1, text opacity=1, draw=white!80!black}
]

\addplot [color=cyan, only marks,  mark=triangle*, mark size=1.5]
  table[row sep=newline]{
1 4.7033    
};
\addlegendentry{Alternating}

\addplot [color=violet, only marks,  mark=o, mark size=1.5]
  table[row sep=newline]{
0.5 2.8948
};
\addlegendentry{Replicated}

\addplot [color=orange_D, only marks,  mark=x, mark size=2]
  table[row sep=newline]{
1 2.7015
};
\addlegendentry{Split}

\addplot [color=red, densely dotted, line width=0.8pt]
  table[row sep=newline]{
0.500000000000000	2.89480000000000
0.510000000000000	2.80100000000000
0.520000000000000	2.71390000000000
0.530000000000000	2.63270000000000
0.540000000000000	2.55690000000000
0.550000000000000	2.48600000000000
0.560000000000000	2.41930000000000
0.570000000000000	2.35660000000000
0.580000000000000	2.29760000000000
0.590000000000000	2.24180000000000
0.600000000000000	2.18900000000000
0.610000000000000	2.13900000000000
0.620000000000000	2.09150000000000
0.630000000000000	2.04630000000000
0.640000000000000	2.00340000000000
0.650000000000000	1.96240000000000
0.660000000000000	1.92330000000000
0.670000000000000	1.88600000000000
0.680000000000000	1.85020000000000
0.690000000000000	1.81600000000000
0.700000000000000	1.78310000000000
0.710000000000000	1.75160000000000
0.720000000000000	1.72140000000000
0.730000000000000	1.69220000000000
0.740000000000000	1.66420000000000
0.750000000000000	1.63720000000000
0.760000000000000	1.61120000000000
0.770000000000000	1.58610000000000
0.780000000000000	1.56190000000000
0.790000000000000	1.53850000000000
0.800000000000000	1.51580000000000
0.810000000000000	1.49390000000000
0.820000000000000	1.47270000000000
0.830000000000000	1.45220000000000
0.840000000000000	1.43230000000000
0.850000000000000	1.41300000000000
0.860000000000000	1.39420000000000
0.870000000000000	1.37600000000000
0.880000000000000	1.35840000000000
0.890000000000000	1.34120000000000
0.900000000000000	1.32450000000000
0.910000000000000	1.30820000000000
0.920000000000000	1.29240000000000
0.930000000000000	1.27700000000000
0.940000000000000	1.26210000000000
0.950000000000000	1.24740000000000
0.960000000000000	1.23320000000000
0.970000000000000	1.21930000000000
0.980000000000000	1.20580000000000
0.990000000000000	1.19250000000000
1	1.17960000000000
  };
\addlegendentry{Coded (LQ)}

\addplot [color=black, densely dashed, line width=0.8pt]
  table[row sep=newline]{
0.500000000000000	6.64620000000000
0.510000000000000	6.43060000000000
0.520000000000000	6.23030000000000
0.530000000000000	6.04360000000000
0.540000000000000	5.86930000000000
0.550000000000000	5.70600000000000
0.560000000000000	5.55270000000000
0.570000000000000	5.40860000000000
0.580000000000000	5.27270000000000
0.590000000000000	5.14440000000000
0.600000000000000	5.02300000000000
0.610000000000000	4.90790000000000
0.620000000000000	4.79870000000000
0.630000000000000	4.69490000000000
0.640000000000000	4.59610000000000
0.650000000000000	4.50190000000000
0.660000000000000	4.41200000000000
0.670000000000000	4.32600000000000
0.680000000000000	4.24380000000000
0.690000000000000	4.16500000000000
0.700000000000000	4.08950000000000
0.710000000000000	4.01710000000000
0.720000000000000	3.94740000000000
0.730000000000000	3.88050000000000
0.740000000000000	3.81600000000000
0.750000000000000	3.75400000000000
0.760000000000000	3.69410000000000
0.770000000000000	3.63640000000000
0.780000000000000	3.58070000000000
0.790000000000000	3.52680000000000
0.800000000000000	3.47470000000000
0.810000000000000	3.42440000000000
0.820000000000000	3.37560000000000
0.830000000000000	3.32830000000000
0.840000000000000	3.28260000000000
0.850000000000000	3.23810000000000
0.860000000000000	3.19500000000000
0.870000000000000	3.15320000000000
0.880000000000000	3.11260000000000
0.890000000000000	3.07310000000000
0.900000000000000	3.03470000000000
0.910000000000000	2.99730000000000
0.920000000000000	2.96090000000000
0.930000000000000	2.92550000000000
0.940000000000000	2.89100000000000
0.950000000000000	2.85740000000000
0.960000000000000	2.82470000000000
0.970000000000000	2.79270000000000
0.980000000000000	2.76160000000000
0.990000000000000	2.73120000000000
1	2.70150000000000
};
\addlegendentry{Coded (HQ)}

\end{axis}
\end{tikzpicture}
        \caption{Inter-frame period $\tau=2$.}
        \label{fig:lat_eta_D2}
    \end{subfigure}	
        	\begin{subfigure}[b]{.49\linewidth}
	    \centering
        \begin{tikzpicture}
\pgfplotsset{every tick label/.append style={font=\scriptsize}}
\tikzstyle{dotted}= [dash pattern=on \pgflinewidth off 0.5mm] 
\tikzstyle{dashed}= [dash pattern=on 7.5*0.8*0.8pt off 7.5*0.4*0.8pt]
\tikzstyle{dashdotted} = [dash pattern=on 7.5*0.8*0.6pt off 7.5*0.8*0.3pt on \the\pgflinewidth off 7.5*0.8*0.3pt]
\tikzstyle{dotted2} = [dash pattern=on 7.5*0.8*0.3pt off 7.5*0.8*0.2pt]

\begin{axis}[%
width=\sfwidth,
height=\sfheight,
at={(0,0)},
scale only axis,
xmin=0.5,
xmax=1,
xlabel near ticks,
xlabel style={font=\footnotesize\color{white!15!black}},
xlabel={$\eta$},
ylabel near ticks,
ymin=0,
ymax=10,
ylabel style={font=\footnotesize\color{white!15!black}},
ylabel={$T  _{99}$},
axis background/.style={fill=white},
xmajorgrids,
ymajorgrids,
legend style={font=\tiny, at={(0.99,0.98)}, anchor=north east, legend cell align=left, align=left, fill opacity=0.8, draw opacity=1, text opacity=1, draw=white!80!black}
]

\addplot [color=cyan, only marks,  mark=triangle*, mark size=1.5]
  table[row sep=newline]{
1 4.6117  
};
\addlegendentry{Alternating}

\addplot [color=violet, only marks,  mark=o, mark size=1.5]
  table[row sep=newline]{
0.5 2.3542
};
\addlegendentry{Replicated}

\addplot [color=orange_D, only marks,  mark=x, mark size=2]
  table[row sep=newline]{
1 2.6488
};
\addlegendentry{Split}

\addplot [color=red, densely dotted, line width=0.8pt]
  table[row sep=newline]{
0.500000000000000	2.35420000000000
0.510000000000000	2.30430000000000
0.520000000000000	2.25670000000000
0.530000000000000	2.21120000000000
0.540000000000000	2.16760000000000
0.550000000000000	2.12590000000000
0.560000000000000	2.08580000000000
0.570000000000000	2.04730000000000
0.580000000000000	2.01030000000000
0.590000000000000	1.97470000000000
0.600000000000000	1.94040000000000
0.610000000000000	1.90740000000000
0.620000000000000	1.87560000000000
0.630000000000000	1.84480000000000
0.640000000000000	1.81510000000000
0.650000000000000	1.78640000000000
0.660000000000000	1.75860000000000
0.670000000000000	1.73170000000000
0.680000000000000	1.70570000000000
0.690000000000000	1.68040000000000
0.700000000000000	1.65600000000000
0.710000000000000	1.63220000000000
0.720000000000000	1.60920000000000
0.730000000000000	1.58680000000000
0.740000000000000	1.56510000000000
0.750000000000000	1.54400000000000
0.760000000000000	1.52340000000000
0.770000000000000	1.50340000000000
0.780000000000000	1.48400000000000
0.790000000000000	1.46500000000000
0.800000000000000	1.44660000000000
0.810000000000000	1.42860000000000
0.820000000000000	1.41110000000000
0.830000000000000	1.39400000000000
0.840000000000000	1.37730000000000
0.850000000000000	1.36100000000000
0.860000000000000	1.34510000000000
0.870000000000000	1.32960000000000
0.880000000000000	1.31450000000000
0.890000000000000	1.29970000000000
0.900000000000000	1.28520000000000
0.910000000000000	1.27110000000000
0.920000000000000	1.25730000000000
0.930000000000000	1.24370000000000
0.940000000000000	1.23050000000000
0.950000000000000	1.21750000000000
0.960000000000000	1.20490000000000
0.970000000000000	1.19250000000000
0.980000000000000	1.18030000000000
0.990000000000000	1.16840000000000
1	1.15670000000000
  };
\addlegendentry{Coded (LQ)}

\addplot [color=black, densely dashed, line width=0.8pt]
  table[row sep=newline]{
0.500000000000000	5.40290000000000
0.510000000000000	5.28830000000000
0.520000000000000	5.17880000000000
0.530000000000000	5.07410000000000
0.540000000000000	4.97380000000000
0.550000000000000	4.87780000000000
0.560000000000000	4.78560000000000
0.570000000000000	4.69710000000000
0.580000000000000	4.61200000000000
0.590000000000000	4.53020000000000
0.600000000000000	4.45130000000000
0.610000000000000	4.37540000000000
0.620000000000000	4.30210000000000
0.630000000000000	4.23140000000000
0.640000000000000	4.16300000000000
0.650000000000000	4.09700000000000
0.660000000000000	4.03310000000000
0.670000000000000	3.97130000000000
0.680000000000000	3.91140000000000
0.690000000000000	3.85330000000000
0.700000000000000	3.79710000000000
0.710000000000000	3.74250000000000
0.720000000000000	3.68950000000000
0.730000000000000	3.63800000000000
0.740000000000000	3.58800000000000
0.750000000000000	3.53940000000000
0.760000000000000	3.49220000000000
0.770000000000000	3.44620000000000
0.780000000000000	3.40150000000000
0.790000000000000	3.35790000000000
0.800000000000000	3.31540000000000
0.810000000000000	3.27410000000000
0.820000000000000	3.23380000000000
0.830000000000000	3.19450000000000
0.840000000000000	3.15610000000000
0.850000000000000	3.11870000000000
0.860000000000000	3.08210000000000
0.870000000000000	3.04650000000000
0.880000000000000	3.01160000000000
0.890000000000000	2.97760000000000
0.900000000000000	2.94430000000000
0.910000000000000	2.91180000000000
0.920000000000000	2.88000000000000
0.930000000000000	2.84890000000000
0.940000000000000	2.81850000000000
0.950000000000000	2.78870000000000
0.960000000000000	2.75950000000000
0.970000000000000	2.73100000000000
0.980000000000000	2.70300000000000
0.990000000000000	2.67560000000000
1	2.64880000000000
};
\addlegendentry{Coded (HQ)}

\end{axis}
\end{tikzpicture}
        \caption{Inter-frame period $\tau=4$.}
        \label{fig:lat_eta_D4}
    \end{subfigure}	
     \caption{99th percentile $T_{99}$ of the latency as a function of the coding rate $\eta$ in an error-free $D/M/2$ with $\mu_1=\mu_2=1$.}
 \label{fig:lat_eta}
\end{figure}
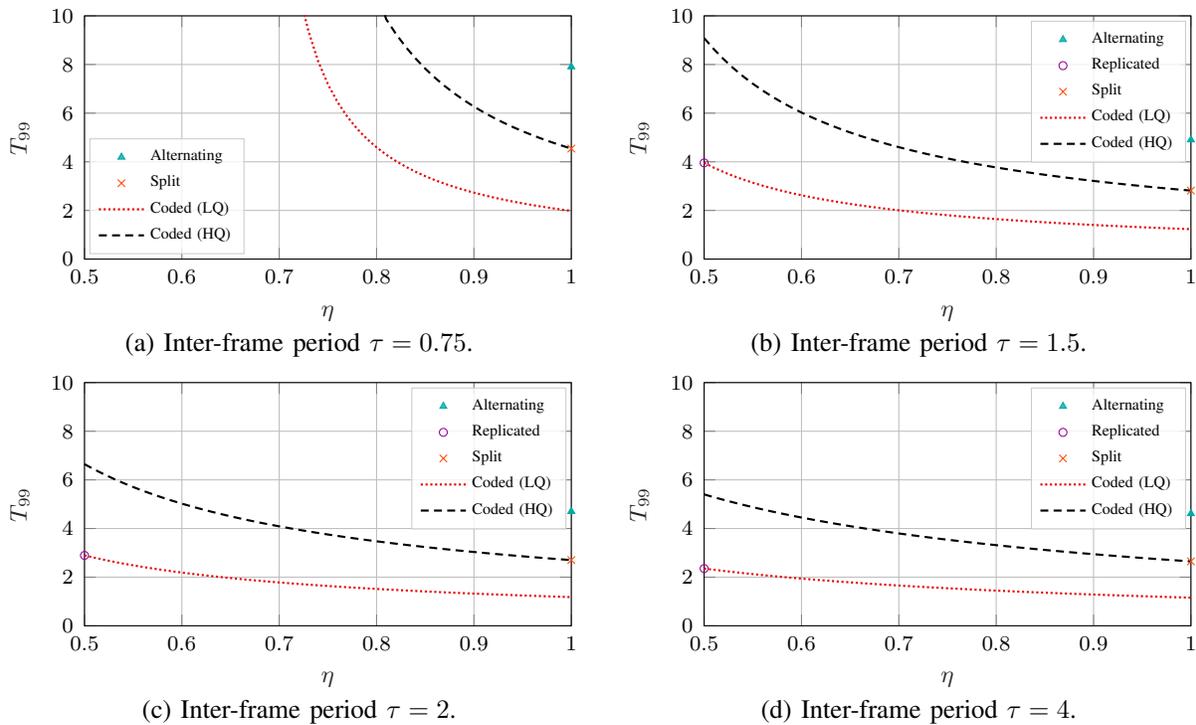

Naturally, the coded system depends on the efficiency of the code: an \gls{mdc} scheme with a higher $\eta$ will have smaller packets, but each individual packet will contain less information on the frame, so the lower-quality version will have a worse quality than the equivalent for a scheme with a smaller $\eta$. The replicated and split scheme are the two extremes: in the replicated scheme, each packet is enough on its own to get the full-quality frame, while in the split scheme, an individual packet is not enough to decode the frame at any quality. Fig.~\ref{fig:lat_eta} shows the worst-case latency performance of the scheme, represented using the 99th percentile of the system time, along with the three other schemes. We have $\eta=1$ for the alternating and split schemes, and $\eta=0.5$ for the replicated scheme. As the plots show, there is an inverse relationship between the latency performance and the quality achievable with just one packet, as increasing the redundancy leads to a sharp increase in the latency, and the system can become unstable if packets are both frequent and large, as is shown in Fig.~\ref{fig:lat_eta_D075}, which has $\tau=0.75$. In this case, the load on the two paths is already high, so choosing a low $\eta$ can increase the queuing delay significantly, with a significant delay increase. If the two paths have a very low load, as in Fig.~\ref{fig:lat_eta_D4}, which has $\tau=4$, the benefits from a more efficient code are smaller, and the latency cost of having a higher quality from a single packet is lower.

\begin{figure}
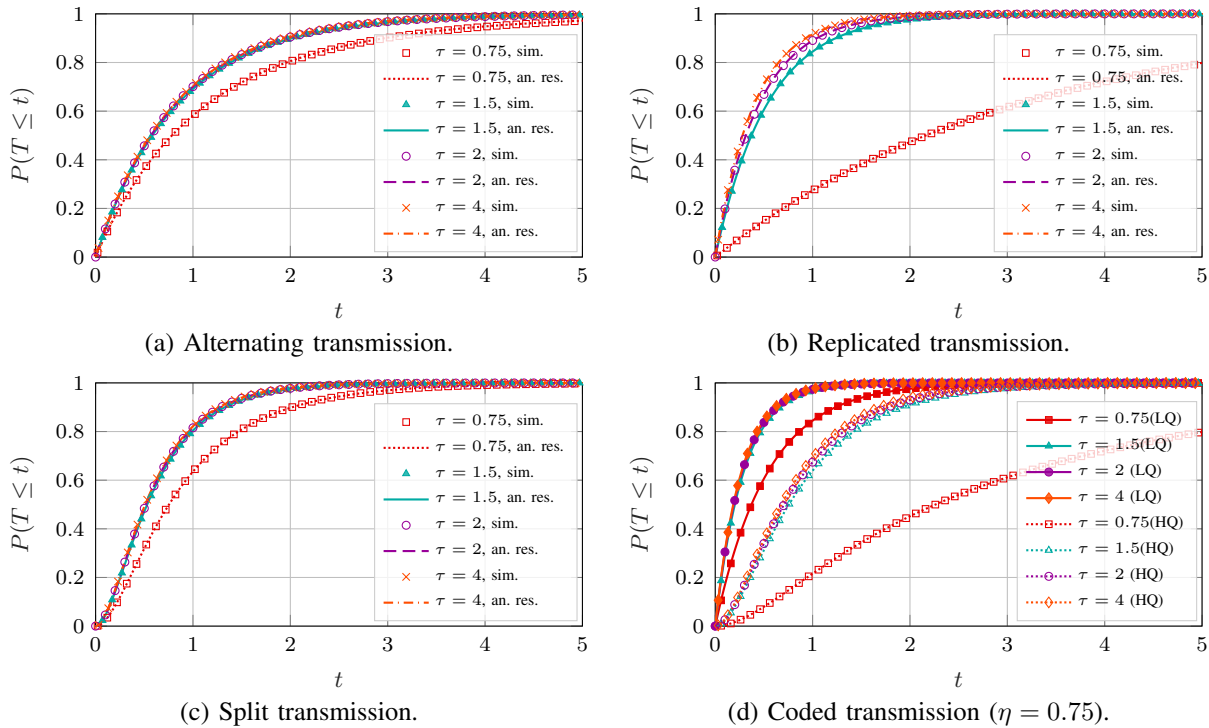

    \centering
	\begin{subfigure}[b]{.49\linewidth}
	    \centering
        \input{./figures/dm2_lat_dist_rho_alt_unb.tex}\vspace{-0.7cm}
        \caption{Alternating transmission.}
        \label{fig:lat_dist_noerr_alt_unb}
    \end{subfigure}	
	\centering
	\begin{subfigure}[b]{.49\linewidth}
	    \centering
        \input{./figures/dm2_lat_dist_rho_min_unb.tex}\vspace{-0.7cm}
        \caption{Replicated transmission.}
        \label{fig:lat_dist_noerr_min_unb}
    \end{subfigure}	
	\begin{subfigure}[b]{.49\linewidth}
	    \centering
        \input{./figures/dm2_lat_dist_rho_max_unb.tex}\vspace{-0.7cm}
        \caption{Split transmission.}
        \label{fig:lat_dist_noerr_max_unb}
    \end{subfigure}	
    	\begin{subfigure}[b]{.49\linewidth}
	    \centering
        \input{./figures/dm2_lat_dist_rho_code_unb.tex}\vspace{-0.7cm}
        \caption{Coded transmission ($\eta=0.75$).}
        \label{fig:lat_dist_noerr_cod_unb}
    \end{subfigure}	
     \caption{Latency \gls{cdf} for the four schemes in an error-free $D/M/2$ with $\mu_1=1$ and $\mu_2=1.5$.}
 \label{fig:lat_CDF_unb}
\end{figure}

We can then look at what happens in an unbalanced system, in which the two paths have different service rates: Fig.~\ref{fig:lat_CDF_unb} shows the \gls{cdf} of the latency for the four schemes in a system with $\mu_1=1$ and $\mu_2=1.5$. In this case, the schemes that can rely on the faster path to transmit the frame, i.e., the replicated and coded schemes, can significantly improve their performance, while the tail of the distribution for the alternating scheme remains similar to the balanced case: packets on the slower path can still have long delays. The same thing happens for the split system, for which the slower path is still a limiting factor: while performance is slightly better than in the balanced scenario, the improvements are far smaller than the one of the replicated system. We note that alternating scheme can benefit from a certain adaptation when the service rates are unbalanced; for example, one possibility could be to try to keep the load on each of the channels equal.

\begin{figure}[!t]
 \centering
 \input{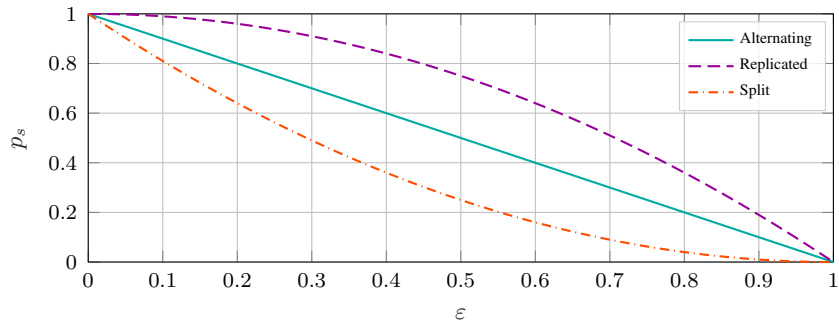}
 \caption{Frame delivery probability as a function of $\varepsilon$ in an error-prone $D/M/2$ with $\varepsilon_1=\varepsilon_2=\varepsilon$.}
 \label{fig:err_prob}
\end{figure}

\begin{figure}[!t]
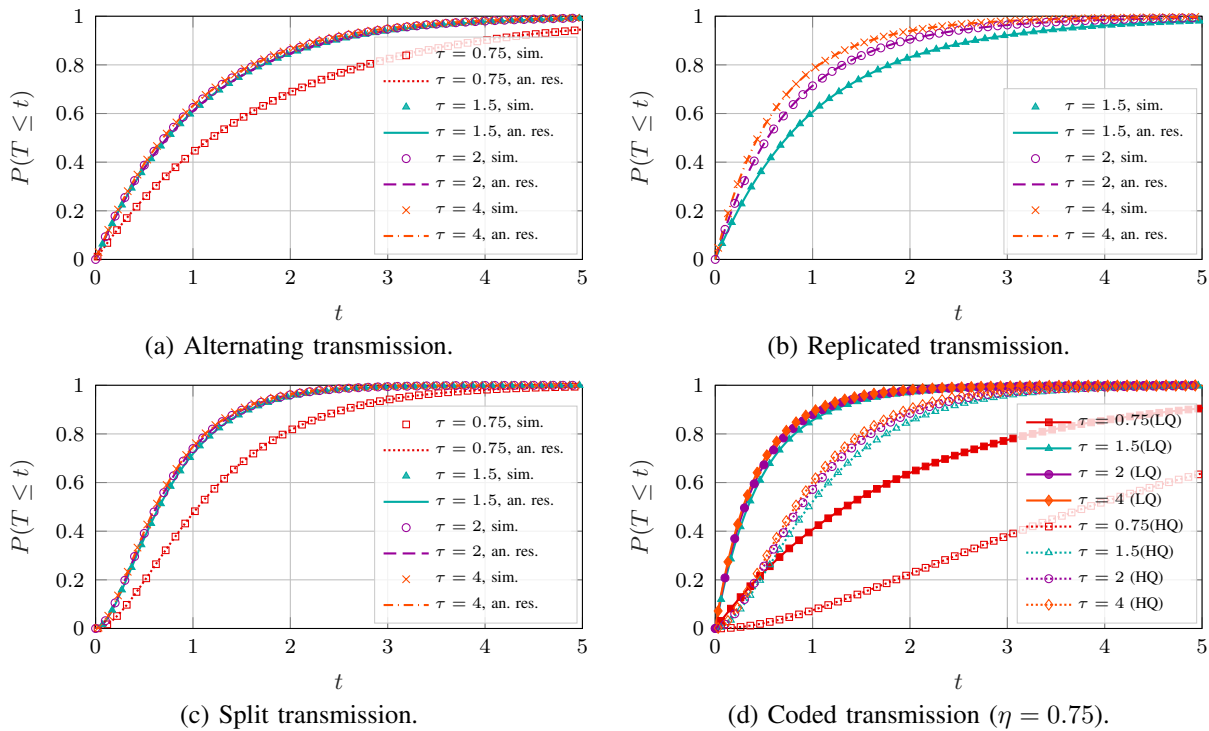

    \centering
	\begin{subfigure}[b]{.49\linewidth}
	    \centering
        \input{./figures/dm2_lat_dist_rho_alt_err.tex}\vspace{-0.7cm}
        \caption{Alternating transmission.}
        \label{fig:lat_dist_err_alt}
    \end{subfigure}	
	\centering
	\begin{subfigure}[b]{.49\linewidth}
	    \centering
        \input{./figures/dm2_lat_dist_rho_min_err.tex}\vspace{-0.7cm}
        \caption{Replicated transmission.}
        \label{fig:lat_dist_err_min}
    \end{subfigure}	
	\begin{subfigure}[b]{.49\linewidth}
	    \centering
        \input{./figures/dm2_lat_dist_rho_max_err.tex}\vspace{-0.7cm}
        \caption{Split transmission.}
        \label{fig:lat_dist_err_max}
    \end{subfigure}	
    	\begin{subfigure}[b]{.49\linewidth}
	    \centering
        \input{./figures/dm2_lat_dist_rho_code_err.tex}\vspace{-0.7cm}
        \caption{Coded transmission ($\eta=0.75$).}
        \label{fig:lat_dist_err_cod}
    \end{subfigure}	
     \caption{Latency \gls{cdf} for the four schemes in an error-prone $D/M/2$ with $\mu_1=\mu_2=1$ and $\varepsilon_1=\varepsilon_2=0.2$.}
 \label{fig:lat_CDF_bal_err}
\end{figure}

Finally, we consider what happens in an error-prone system, as the transmission schemes provide different protection from errors. While the replicated scheme provides extra redundancy by duplicating the packet, the alternating one provides no protection at all against packet erasures, while the split scheme does even worse. In order for a frame to be decoded in the split scheme, both halves need to be delivered to the receiver. Fig.~\ref{fig:err_prob} shows the frame decoding probability as a function of the packet erasure probability (considering the same erasure probability on both paths): as expected, the replicated scheme is much more reliable, and the split scheme is less so, while the alternating scheme follows the erasure probability exactly. In the coded scheme, the probability of decoding a frame at any quality is the same as for the replicated scheme, while the probability of getting a full-quality frame follows the curve for the split system, as it requires both path to deliver their part correctly.

\begin{figure}
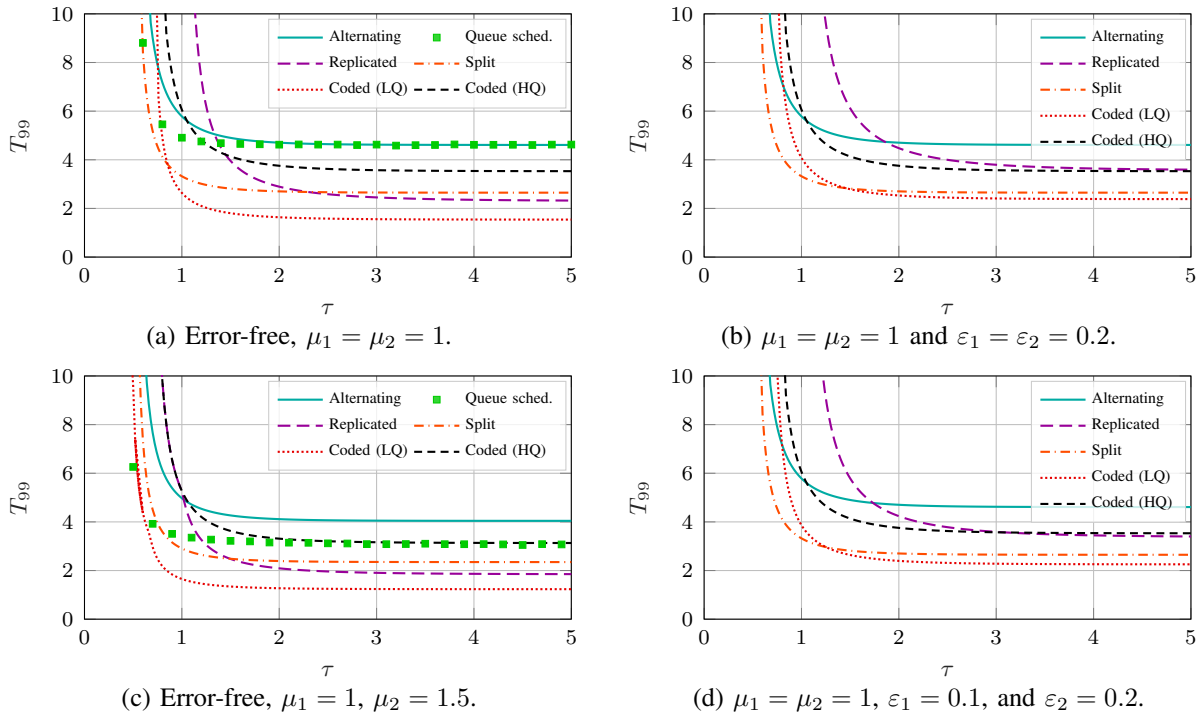

    \centering
	\begin{subfigure}[b]{.49\linewidth}
        \centering
        \input{./figures/dm2_lat99_D.tex}
        \caption{Error-free, $\mu_1=\mu_2=1$.}
        \label{fig:lat_D_noerr}
    \end{subfigure}	
	\centering
	\begin{subfigure}[b]{.49\linewidth}
        \centering
        \input{./figures/dm2_lat99_D_err.tex}
        \caption{$\mu_1=\mu_2=1$ and $\varepsilon_1=\varepsilon_2=0.2$.}
        \label{fig:lat_D_err}
    \end{subfigure}	
	\begin{subfigure}[b]{.49\linewidth}
        \centering
        \input{./figures/dm2_lat99_D_unb.tex}
        \caption{Error-free, $\mu_1=1$, $\mu_2=1.5$.}
        \label{fig:lat_D_unb}
    \end{subfigure}	
    \begin{subfigure}[b]{.49\linewidth}
        \centering
        \input{./figures/dm2_lat99_D_unberr.tex}
        \caption{$\mu_1=\mu_2=1$, $\varepsilon_1=0.1$, and $\varepsilon_2=0.2$.}
        \label{fig:lat_D_unberr}
    \end{subfigure}	
     \caption{99th percentile $T_{99}$ of the latency as a function of the inter-frame time in a $D/M/2$. The coded transmission has $\eta=0.75$.}
 \label{fig:lat_D}
\end{figure}

This difference between the schemes is reflected in the \gls{cdf} of the latency of successfully delivered frames, which is shown in Fig.~\ref{fig:lat_CDF_bal_err} for a system with $\mu_1=\mu_2=1$ and $\varepsilon_1=\varepsilon_2=0.2$. In this case, the distributions for the alternating and split scheme, as well as for the high-quality version of the coded transmission, are exactly the same as in Fig.~\ref{fig:lat_CDF_bal}. This is because any packet loss causes the whole frame to be lost, and is then not computed in the latency distribution. On the other hand, the replicated scheme and the low-quality version of the coded transmission have a higher delay: this is because they can recover from errors on one path by using the other, but cannot then use the benefits of path diversity to improve latency. In general, the far lower error rate of these schemes makes them preferable in systems with a high packet erasure probability.

We can also look at system optimization, if we have an application that can work at different frame rates: Fig.~\ref{fig:lat_D} shows the 99th percentile of the latency as a function of $\tau$ for 4 different systems. As expected, a longer inter-frame time leads to a reduced latency in all cases: as the load on the system decreases, the probability of having to wait for the queued frames to be sent becomes smaller. The alternating system, which does not exploit the parallel paths fully, can support a high frame rate but performs worse than the other systems when the load is low (the 99th percentile of the service time, with $\mu=1$, would be equal to $-\log(0.01)\simeq4.6$). On the other hand, the replicated system has a very high load, and underperforms for low values of $\tau$, but can actually keep a low latency when the load on the system is low. The split system provides a middle ground, but it is also highly vulnerable to errors, as previously discussed. Finally, the \gls{mdc} solution can provide a very good delay, smaller than all the other schemes', for the low-quality version, while still delivering the full-quality frame in a reasonable time. We also provide the simulation results for a queue-based scheduler, which transmits the packet over the path with the shortest queue, in the two error-free scenarios. As expected, a pure round-robin scheduler like the alternating scheme underperforms in the unbalanced scenario, but it actually performs very well in the balanced one, except for very high frame rates. This is because the two paths will have a similar queue most of the time, so the alternating scheme is close to the optimal uncoded scheme if the load is relatively low.

\subsection{Peak Age of Information}

\begin{figure}
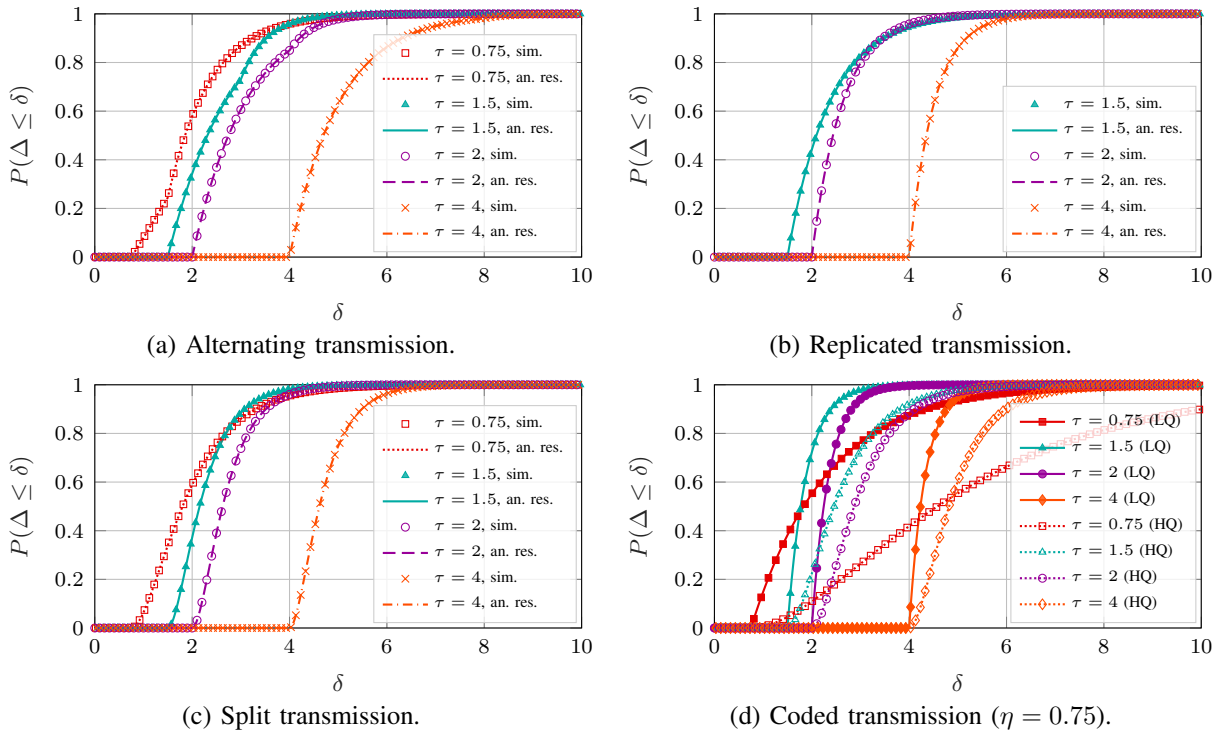

    \centering
	\begin{subfigure}[b]{.49\linewidth}
	    \centering
        \input{./figures/dm2_paoi_dist_rho_alt.tex}\vspace{-0.7cm}
        \caption{Alternating transmission.}
        \label{fig:paoi_dist_noerr_alt}
    \end{subfigure}	
	\centering
	\begin{subfigure}[b]{.49\linewidth}
	    \centering
        \input{./figures/dm2_paoi_dist_rho_min.tex}\vspace{-0.7cm}
        \caption{Replicated transmission.}
        \label{fig:paoi_dist_noerr_min}
    \end{subfigure}	
	\begin{subfigure}[b]{.49\linewidth}
	    \centering
        \input{./figures/dm2_paoi_dist_rho_max.tex}\vspace{-0.7cm}
        \caption{Split transmission.}
        \label{fig:paoi_dist_noerr_max}
    \end{subfigure}	
        	\begin{subfigure}[b]{.49\linewidth}
	    \centering
        \input{./figures/dm2_paoi_dist_rho_code.tex}\vspace{-0.7cm}
        \caption{Coded transmission ($\eta=0.75$).}
        \label{fig:paoi_dist_noerr_cod}
    \end{subfigure}	
     \caption{\gls{paoi} distribution for the three schemes in an error-free $D/M/2$ with $\mu_1=\mu_2=1$. The coded transmission has $\eta=0.75$.}
 \label{fig:paoi_CDF_bal}
\end{figure}

We now analyze the \gls{paoi} performance of the four schemes. We first look at the simple error-free, balanced system with two paths with the same service rate. Fig.~\ref{fig:paoi_CDF_bal} shows the \gls{cdf} of the \gls{paoi} for the four schemes. Unlike the system time, \gls{paoi} does not necessarily benefit from a lower load on the system, as it also includes the inter-frame time: the scenario with $\tau=4$ has the worst \gls{paoi} for all four schemes. It is interesting to note that the \gls{cdf} for the alternating scheme, shown in Fig.~\ref{fig:paoi_dist_noerr_alt}, is not smooth: this is because in some cases frame $i$ arrives before frame $i-1$, which is transmitted on the other path. This out-of-order arrival can only result in a \gls{paoi} of at least $2\tau$, as the age of the $i-2$-th frame, which has certainly arrived before the $i$-th as it was sent on the same path, is $2\tau$ when the $i$-th frame is sent. Interestingly, the \glspl{cdf} for different values of $\tau$ also cross, which is an indication that systems optimized for the average \gls{paoi} will not have optimal worst-case performance. This is also true for the replicated and split schemes, which have similar \gls{paoi} distributions, as shown in Fig.~\ref{fig:paoi_dist_noerr_min} and Fig.~\ref{fig:paoi_dist_noerr_max}. As for the latency, the coded scheme, shown in Fig.~\ref{fig:paoi_dist_noerr_cod}, can outperform all others and deliver the lower-quality version with a significantly lower \gls{paoi}, while the full-quality version is somewhat slower in most cases. The coded scheme also has a case with very high load, for $\tau=0.75$, which is close to its stability limit: in this case, the load on the system is too high and the \gls{paoi} performance is not good. This further highlights the need to optimize the frame rate, as setting the wrong frame rate can significantly increase the \gls{paoi}, consequently decreasing the \gls{qoe} of the \gls{vr} service.

\begin{figure}
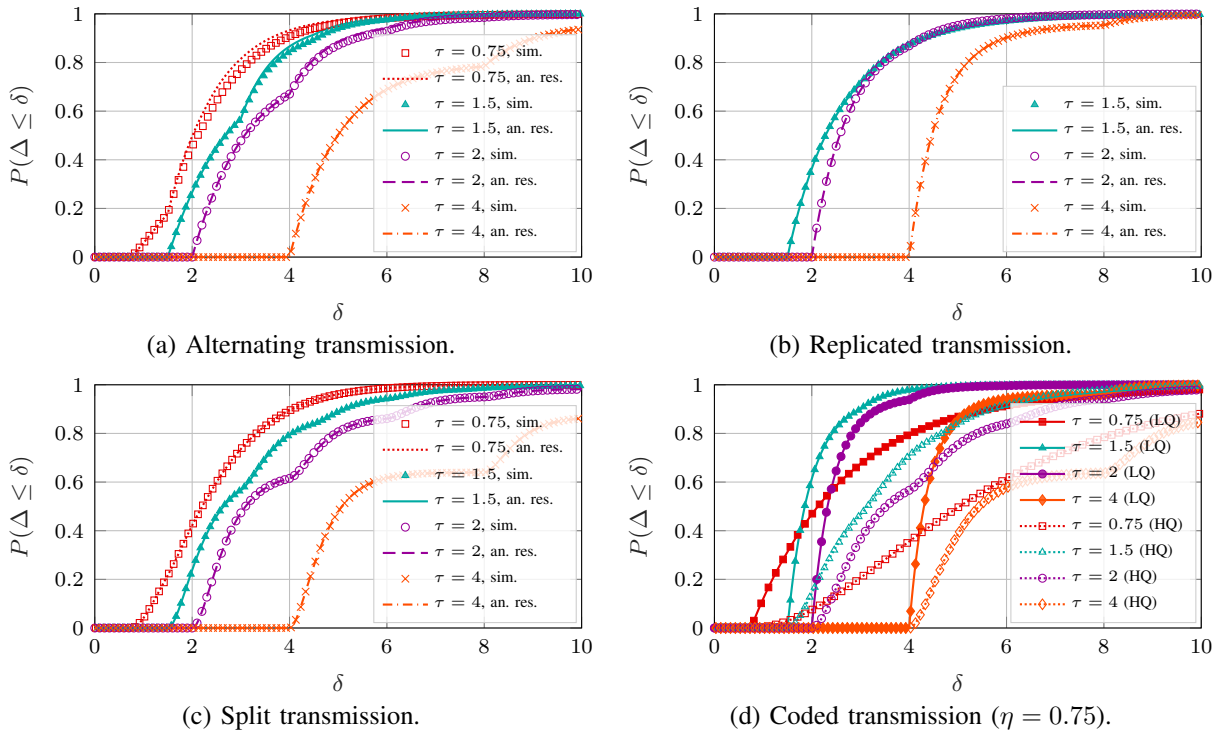

    \centering
	\begin{subfigure}[b]{.49\linewidth}
	    \centering
        \input{./figures/dm2_paoi_dist_rho_alt_err.tex}\vspace{-0.7cm}
        \caption{Alternating transmission.}
        \label{fig:paoi_dist_err_alt}
    \end{subfigure}	
	\centering
	\begin{subfigure}[b]{.49\linewidth}
	    \centering
        \input{./figures/dm2_paoi_dist_rho_min_err.tex}\vspace{-0.7cm}
        \caption{Replicated transmission.}
        \label{fig:paoi_dist_err_min}
    \end{subfigure}	
	\begin{subfigure}[b]{.49\linewidth}
	    \centering
        \input{./figures/dm2_paoi_dist_rho_max_err.tex}\vspace{-0.7cm}
        \caption{Split transmission.}
        \label{fig:paoi_dist_err_max}
    \end{subfigure}	
        	\begin{subfigure}[b]{.49\linewidth}
	    \centering
        \input{./figures/dm2_paoi_dist_rho_code_err.tex}\vspace{-0.7cm}
        \caption{Coded transmission ($\eta=0.75$).}
        \label{fig:paoi_dist_err_cod}
    \end{subfigure}	
     \caption{\gls{paoi} distribution for the three schemes in an error-prone $D/M/2$ with $\mu_1=\mu_2=1$ and $\varepsilon_1=\varepsilon_2=0.2$. The coded transmission has $\eta=0.75$.}
 \label{fig:paoi_CDF_bal_err}
\end{figure}

As we discussed in Sec.~\ref{sec:an_alt}, we do not compute the precise \gls{cdf} of the \gls{paoi} of the alternating scheme in a scenario with errors, as we do not consider the possibility of multiple out-of-order packets, which can further increase the age. Fig.~\ref{fig:paoi_CDF_bal_err}, which shows the \gls{cdf} of the peak age in an error-prone system in which both channels have an error rate of $0.2$, shows this: in Fig.~\ref{fig:paoi_dist_err_alt}, the actual age is higher than the bound. The bound is tighter for low system loads, and there are now multiple flection points that correspond to successive errors in the transmission. The errors also cause flection points in Fig.~\ref{fig:paoi_dist_err_min}, and the tail of the age is shifted to the right because of packet loss. This phenomenon is much worse for the split scheme, as shown in Fig.~\ref{fig:paoi_dist_err_max}, as the split system is much more vulnerable to errors. This can cause the \gls{paoi} to increase significantly. The same is true for the full-quality \gls{paoi} of the coded scheme, as depicted in Fig.~\ref{fig:paoi_dist_err_cod}, while the lower-quality version of the frames can actually be delivered with a relatively low \gls{paoi} even with a high error rate.

\begin{figure}
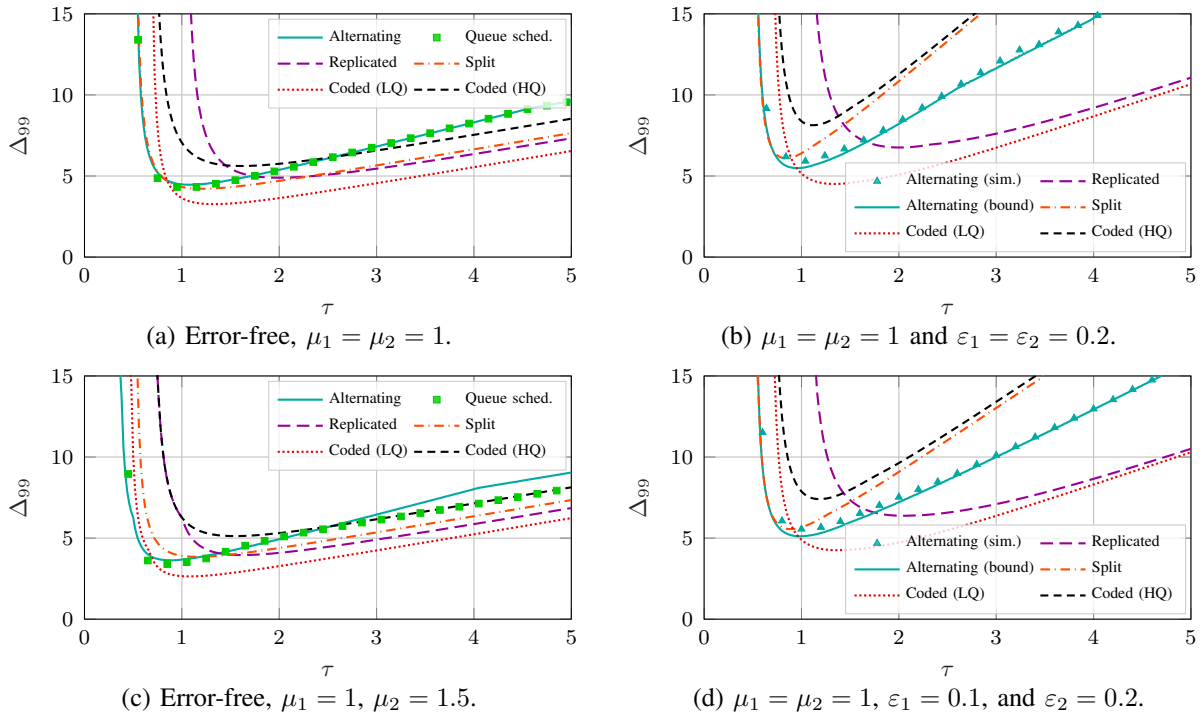

    \centering
	\begin{subfigure}[b]{.49\linewidth}
        \centering
        \input{./figures/dm2_paoi_rho.tex}
        \caption{Error-free, $\mu_1=\mu_2=1$.}
        \label{fig:paoi_D_noerr}
    \end{subfigure}	
	\centering
	\begin{subfigure}[b]{.49\linewidth}
        \centering
        \input{./figures/dm2_paoi_rho_err.tex}
        \caption{$\mu_1=\mu_2=1$ and $\varepsilon_1=\varepsilon_2=0.2$.}
        \label{fig:paoi_D_err}
    \end{subfigure}	
	\begin{subfigure}[b]{.49\linewidth}
        \centering
        \input{./figures/dm2_paoi_rho_unb.tex}
        \caption{Error-free, $\mu_1=1$, $\mu_2=1.5$.}
        \label{fig:paoi_D_unb}
    \end{subfigure}	
    \begin{subfigure}[b]{.49\linewidth}
        \centering
        \input{./figures/dm2_paoi_rho_unberr.tex}
        \caption{$\mu_1=\mu_2=1$, $\varepsilon_1=0.1$, and $\varepsilon_2=0.2$.}
        \label{fig:paoi_D_unberr}
    \end{subfigure}	
     \caption{99th percentile $\Delta_{99}$ of the \gls{paoi} as a function of the inter-frame time in a $D/M/2$. The coded transmission has $\eta=0.75$.}
 \label{fig:paoi_D}
\end{figure}

We now look at the optimization of the system, using the 99th percentile of the \gls{paoi} as a metric for the worst-case performance. Fig.~\ref{fig:paoi_D} shows what happens as we change $\tau$ for the four considered schemes. All results have a U shape, with an optimal frame rate which balances the inter-frame period and the latency of frames that are sent. Naturally, this optimal point is different for different schemes and scenarios. However, it is interesting to note that the lower-quality version of the coded scheme always has the lowest age, aside from having the lowest frame loss. On the other hand, the replicated scheme always performs slightly worse than the split scheme, as even though it will lose fewer frames, its higher load significantly increases the optimal inter-frame period. We can also see that the real age for the alternating system is close to the bound for the plots with error (shown on the right). The alternating scheme can perform slightly better than the split scheme, but is outperformed by the coded scheme at the lower quality. On the other hand, the full-quality version of the coded frames has a very high age, as it has no protection from errors and the higher load associated to the redundancy in the \gls{mdc} scheme. In general, the alternating scheme performs surprisingly well in terms of \gls{paoi}. All the schemes could benefit from a more advanced scheduler, regulating the load on the two paths based on the number of packets still in the queue: unlike latency-oriented systems, which should have a very low load (and, consequently, little effect from improving the scheduler, as the queues are almost always empty), \gls{paoi} is minimized for rather high load values, close to 0.5, so having an intelligent scheduler can be a significant improvement.

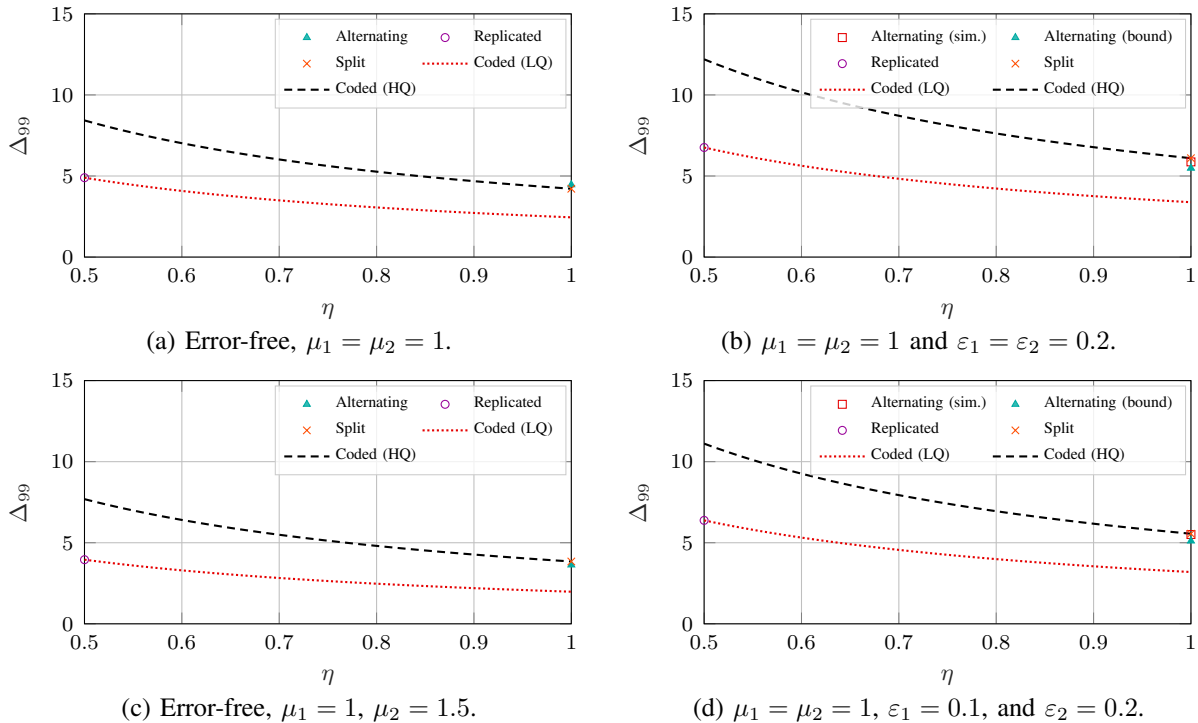
\begin{figure}
    \centering
	\begin{subfigure}[b]{.49\linewidth}
        \centering
        \begin{tikzpicture}
\pgfplotsset{every tick label/.append style={font=\scriptsize}}
\tikzstyle{dotted}= [dash pattern=on \pgflinewidth off 0.5mm] 
\tikzstyle{dashed}= [dash pattern=on 7.5*0.8*0.8pt off 7.5*0.4*0.8pt]
\tikzstyle{dashdotted} = [dash pattern=on 7.5*0.8*0.6pt off 7.5*0.8*0.3pt on \the\pgflinewidth off 7.5*0.8*0.3pt]
\tikzstyle{dotted2} = [dash pattern=on 7.5*0.8*0.3pt off 7.5*0.8*0.2pt]

\begin{axis}[%
width=\sfwidth,
height=\sfheight,
at={(0,0)},
scale only axis,
xmin=0.5,
xmax=1,
xlabel near ticks,
xlabel style={font=\footnotesize\color{white!15!black}},
xlabel={$\eta$},
ylabel near ticks,
ymin=0,
ymax=15,
ylabel style={font=\footnotesize\color{white!15!black}},
ylabel={$\Delta_{99}$},
axis background/.style={fill=white},
xmajorgrids,
ymajorgrids,
legend style={font=\tiny, at={(0.99,0.98)}, anchor=north east,legend columns=2, legend cell align=left, align=left, fill opacity=0.8, draw opacity=1, text opacity=1, draw=white!80!black}
]

\addplot [color=cyan, only marks,  mark=triangle*, mark size=1.5]
  table[row sep=newline]{
1 4.4611
};
\addlegendentry{Alternating}

\addplot [color=violet, only marks,  mark=o, mark size=1.5]
  table[row sep=newline]{
0.5 4.8948
};
\addlegendentry{Replicated}

\addplot [color=orange_D, only marks,  mark=x, mark size=2]
  table[row sep=newline]{
1 4.2139
};
\addlegendentry{Split}

\addplot [color=red, densely dotted, line width=0.8pt]
  table[row sep=newline]{
 0.500000000000000	4.89480000000000
0.510000000000000	4.79890000000000
0.520000000000000	4.70670000000000
0.530000000000000	4.61800000000000
0.540000000000000	4.53260000000000
0.550000000000000	4.45030000000000
0.560000000000000	4.37100000000000
0.570000000000000	4.29430000000000
0.580000000000000	4.22040000000000
0.590000000000000	4.14900000000000
0.600000000000000	4.07980000000000
0.610000000000000	4.01310000000000
0.620000000000000	3.94850000000000
0.630000000000000	3.88590000000000
0.640000000000000	3.82510000000000
0.650000000000000	3.76640000000000
0.660000000000000	3.70940000000000
0.670000000000000	3.65420000000000
0.680000000000000	3.60040000000000
0.690000000000000	3.54830000000000
0.700000000000000	3.49770000000000
0.710000000000000	3.44860000000000
0.720000000000000	3.40070000000000
0.730000000000000	3.35420000000000
0.740000000000000	3.30900000000000
0.750000000000000	3.26490000000000
0.760000000000000	3.22200000000000
0.770000000000000	3.18030000000000
0.780000000000000	3.13950000000000
0.790000000000000	3.09990000000000
0.800000000000000	3.06110000000000
0.810000000000000	3.02340000000000
0.820000000000000	2.98670000000000
0.830000000000000	2.95070000000000
0.840000000000000	2.91560000000000
0.850000000000000	2.88140000000000
0.860000000000000	2.84790000000000
0.870000000000000	2.81530000000000
0.880000000000000	2.78330000000000
0.890000000000000	2.75210000000000
0.900000000000000	2.72170000000000
0.910000000000000	2.69180000000000
0.920000000000000	2.66260000000000
0.930000000000000	2.63400000000000
0.940000000000000	2.60600000000000
0.950000000000000	2.57870000000000
0.960000000000000	2.55180000000000
0.970000000000000	2.52560000000000
0.980000000000000	2.49990000000000
0.990000000000000	2.47470000000000
1	2.44990000000000
  };
\addlegendentry{Coded (LQ)}

\addplot [color=black, densely dashed, line width=0.8pt]
  table[row sep=newline]{
 0.500000000000000	8.42770000000000
0.510000000000000	8.26250000000000
0.520000000000000	8.10360000000000
0.530000000000000	7.95060000000000
0.540000000000000	7.80340000000000
0.550000000000000	7.66150000000000
0.560000000000000	7.52480000000000
0.570000000000000	7.39270000000000
0.580000000000000	7.26530000000000
0.590000000000000	7.14210000000000
0.600000000000000	7.02310000000000
0.610000000000000	6.90800000000000
0.620000000000000	6.79650000000000
0.630000000000000	6.68870000000000
0.640000000000000	6.58410000000000
0.650000000000000	6.48280000000000
0.660000000000000	6.38460000000000
0.670000000000000	6.28930000000000
0.680000000000000	6.19690000000000
0.690000000000000	6.10710000000000
0.700000000000000	6.01980000000000
0.710000000000000	5.93500000000000
0.720000000000000	5.85260000000000
0.730000000000000	5.77240000000000
0.740000000000000	5.69440000000000
0.750000000000000	5.61850000000000
0.760000000000000	5.54460000000000
0.770000000000000	5.47250000000000
0.780000000000000	5.40240000000000
0.790000000000000	5.33400000000000
0.800000000000000	5.26730000000000
0.810000000000000	5.20230000000000
0.820000000000000	5.13890000000000
0.830000000000000	5.07690000000000
0.840000000000000	5.01660000000000
0.850000000000000	4.95750000000000
0.860000000000000	4.89990000000000
0.870000000000000	4.84350000000000
0.880000000000000	4.78850000000000
0.890000000000000	4.73470000000000
0.900000000000000	4.68210000000000
0.910000000000000	4.63070000000000
0.920000000000000	4.58030000000000
0.930000000000000	4.53110000000000
0.940000000000000	4.48280000000000
0.950000000000000	4.43560000000000
0.960000000000000	4.38950000000000
0.970000000000000	4.34420000000000
0.980000000000000	4.29990000000000
0.990000000000000	4.25650000000000
1	4.21390000000000
};
\addlegendentry{Coded (HQ)}

\end{axis}
\end{tikzpicture}
        \caption{Error-free, $\mu_1=\mu_2=1$.}
        \label{fig:paoi_eta_noerr}
    \end{subfigure}	
	\centering
	\begin{subfigure}[b]{.49\linewidth}
        \centering
        \begin{tikzpicture}
\pgfplotsset{every tick label/.append style={font=\scriptsize}}
\tikzstyle{dotted}= [dash pattern=on \pgflinewidth off 0.5mm] 
\tikzstyle{dashed}= [dash pattern=on 7.5*0.8*0.8pt off 7.5*0.4*0.8pt]
\tikzstyle{dashdotted} = [dash pattern=on 7.5*0.8*0.6pt off 7.5*0.8*0.3pt on \the\pgflinewidth off 7.5*0.8*0.3pt]
\tikzstyle{dotted2} = [dash pattern=on 7.5*0.8*0.3pt off 7.5*0.8*0.2pt]

\begin{axis}[%
width=\sfwidth,
height=\sfheight,
at={(0,0)},
scale only axis,
xmin=0.5,
xmax=1,
xlabel near ticks,
xlabel style={font=\footnotesize\color{white!15!black}},
xlabel={$\eta$},
ylabel near ticks,
ymin=0,
ymax=15,
ylabel style={font=\footnotesize\color{white!15!black}},
ylabel={$\Delta_{99}$},
axis background/.style={fill=white},
xmajorgrids,
ymajorgrids,
legend style={font=\tiny, at={(0.99,0.98)}, anchor=north east,legend columns=2, legend cell align=left, align=left, fill opacity=0.8, draw opacity=1, text opacity=1, draw=white!80!black}
]

\addplot [color=red, only marks,  mark=square, mark size=1.5]
  table[row sep=newline]{
1 5.8704
};
\addlegendentry{Alternating (sim.)}

\addplot [color=cyan, only marks,  mark=triangle*, mark size=1.5]
  table[row sep=newline]{
1 5.48210000000000
};
\addlegendentry{Alternating (bound)}

\addplot [color=violet, only marks,  mark=o, mark size=1.5]
  table[row sep=newline]{
0.5 6.7582  
};
\addlegendentry{Replicated}

\addplot [color=orange_D, only marks,  mark=x, mark size=2]
  table[row sep=newline]{
1 6.0987
};
\addlegendentry{Split}

\addplot [color=red, densely dotted, line width=0.8pt]
  table[row sep=newline]{
0.500000000000000	6.75820000000000
0.510000000000000	6.62580000000000
0.520000000000000	6.49860000000000
0.530000000000000	6.37610000000000
0.540000000000000	6.25810000000000
0.550000000000000	6.14440000000000
0.560000000000000	6.03500000000000
0.570000000000000	5.92910000000000
0.580000000000000	5.82720000000000
0.590000000000000	5.72840000000000
0.600000000000000	5.63320000000000
0.610000000000000	5.54080000000000
0.620000000000000	5.45160000000000
0.630000000000000	5.36530000000000
0.640000000000000	5.28160000000000
0.650000000000000	5.20030000000000
0.660000000000000	5.12160000000000
0.670000000000000	5.04530000000000
0.680000000000000	4.97130000000000
0.690000000000000	4.89940000000000
0.700000000000000	4.82950000000000
0.710000000000000	4.76160000000000
0.720000000000000	4.69560000000000
0.730000000000000	4.63130000000000
0.740000000000000	4.56880000000000
0.750000000000000	4.50800000000000
0.760000000000000	4.44870000000000
0.770000000000000	4.39110000000000
0.780000000000000	4.33500000000000
0.790000000000000	4.28040000000000
0.800000000000000	4.22680000000000
0.810000000000000	4.17460000000000
0.820000000000000	4.12440000000000
0.830000000000000	4.07500000000000
0.840000000000000	4.02640000000000
0.850000000000000	3.97900000000000
0.860000000000000	3.93250000000000
0.870000000000000	3.88740000000000
0.880000000000000	3.84310000000000
0.890000000000000	3.80030000000000
0.900000000000000	3.75800000000000
0.910000000000000	3.71670000000000
0.920000000000000	3.67670000000000
0.930000000000000	3.63690000000000
0.940000000000000	3.59860000000000
0.950000000000000	3.56060000000000
0.960000000000000	3.52360000000000
0.970000000000000	3.48740000000000
0.980000000000000	3.45180000000000
0.990000000000000	3.41720000000000
1	3.38290000000000
  };
\addlegendentry{Coded (LQ)}

\addplot [color=black, densely dashed, line width=0.8pt]
  table[row sep=newline]{
0.500000000000000	12.1973000000000
0.510000000000000	11.9577000000000
0.520000000000000	11.7283000000000
0.530000000000000	11.5065000000000
0.540000000000000	11.2940000000000
0.550000000000000	11.0880000000000
0.560000000000000	10.8904000000000
0.570000000000000	10.6995000000000
0.580000000000000	10.5145000000000
0.590000000000000	10.3363000000000
0.600000000000000	10.1644000000000
0.610000000000000	9.99820000000000
0.620000000000000	9.83700000000000
0.630000000000000	9.68060000000000
0.640000000000000	9.52920000000000
0.650000000000000	9.38260000000000
0.660000000000000	9.24060000000000
0.670000000000000	9.10290000000000
0.680000000000000	8.96940000000000
0.690000000000000	8.84200000000000
0.700000000000000	8.71520000000000
0.710000000000000	8.59220000000000
0.720000000000000	8.47010000000000
0.730000000000000	8.35440000000000
0.740000000000000	8.24230000000000
0.750000000000000	8.13150000000000
0.760000000000000	8.02420000000000
0.770000000000000	7.92050000000000
0.780000000000000	7.81930000000000
0.790000000000000	7.71950000000000
0.800000000000000	7.62340000000000
0.810000000000000	7.52970000000000
0.820000000000000	7.43710000000000
0.830000000000000	7.34830000000000
0.840000000000000	7.26030000000000
0.850000000000000	7.17480000000000
0.860000000000000	7.09200000000000
0.870000000000000	7.00970000000000
0.880000000000000	6.93130000000000
0.890000000000000	6.85210000000000
0.900000000000000	6.77730000000000
0.910000000000000	6.70160000000000
0.920000000000000	6.63000000000000
0.930000000000000	6.55740000000000
0.940000000000000	6.48920000000000
0.950000000000000	6.41940000000000
0.960000000000000	6.35370000000000
0.970000000000000	6.28710000000000
0.980000000000000	6.22360000000000
0.990000000000000	6.16040000000000
1	6.09870000000000
};
\addlegendentry{Coded (HQ)}

\end{axis}
\end{tikzpicture}
        \caption{$\mu_1=\mu_2=1$ and $\varepsilon_1=\varepsilon_2=0.2$.}
        \label{fig:paoi_eta_err}
    \end{subfigure}	
	\begin{subfigure}[b]{.49\linewidth}
        \centering
        \begin{tikzpicture}
\pgfplotsset{every tick label/.append style={font=\scriptsize}}
\tikzstyle{dotted}= [dash pattern=on \pgflinewidth off 0.5mm] 
\tikzstyle{dashed}= [dash pattern=on 7.5*0.8*0.8pt off 7.5*0.4*0.8pt]
\tikzstyle{dashdotted} = [dash pattern=on 7.5*0.8*0.6pt off 7.5*0.8*0.3pt on \the\pgflinewidth off 7.5*0.8*0.3pt]
\tikzstyle{dotted2} = [dash pattern=on 7.5*0.8*0.3pt off 7.5*0.8*0.2pt]

\begin{axis}[%
width=\sfwidth,
height=\sfheight,
at={(0,0)},
scale only axis,
xmin=0.5,
xmax=1,
xlabel near ticks,
xlabel style={font=\footnotesize\color{white!15!black}},
xlabel={$\eta$},
ylabel near ticks,
ymin=0,
ymax=15,
ylabel style={font=\footnotesize\color{white!15!black}},
ylabel={$\Delta_{99}$},
axis background/.style={fill=white},
xmajorgrids,
ymajorgrids,
legend style={font=\tiny, at={(0.99,0.98)}, anchor=north east,legend columns=2, legend cell align=left, align=left, fill opacity=0.8, draw opacity=1, text opacity=1, draw=white!80!black}
]

\addplot [color=cyan, only marks,  mark=triangle*, mark size=1.5]
  table[row sep=newline]{
1 3.6303
};
\addlegendentry{Alternating}

\addplot [color=violet, only marks,  mark=o, mark size=1.5]
  table[row sep=newline]{
0.5 3.9542
};
\addlegendentry{Replicated}

\addplot [color=orange_D, only marks,  mark=x, mark size=2]
  table[row sep=newline]{
1 3.844
};
\addlegendentry{Split}

\addplot [color=red, densely dotted, line width=0.8pt]
  table[row sep=newline]{
 0.500000000000000	3.95430000000000
0.510000000000000	3.87690000000000
0.520000000000000	3.80240000000000
0.530000000000000	3.73080000000000
0.540000000000000	3.66180000000000
0.550000000000000	3.59540000000000
0.560000000000000	3.53150000000000
0.570000000000000	3.46930000000000
0.580000000000000	3.40960000000000
0.590000000000000	3.35200000000000
0.600000000000000	3.29620000000000
0.610000000000000	3.24220000000000
0.620000000000000	3.19000000000000
0.630000000000000	3.13940000000000
0.640000000000000	3.09040000000000
0.650000000000000	3.04300000000000
0.660000000000000	2.99700000000000
0.670000000000000	2.95240000000000
0.680000000000000	2.90910000000000
0.690000000000000	2.86690000000000
0.700000000000000	2.82600000000000
0.710000000000000	2.78620000000000
0.720000000000000	2.74760000000000
0.730000000000000	2.71020000000000
0.740000000000000	2.67360000000000
0.750000000000000	2.63790000000000
0.760000000000000	2.60330000000000
0.770000000000000	2.56970000000000
0.780000000000000	2.53660000000000
0.790000000000000	2.50460000000000
0.800000000000000	2.47350000000000
0.810000000000000	2.44290000000000
0.820000000000000	2.41320000000000
0.830000000000000	2.38420000000000
0.840000000000000	2.35580000000000
0.850000000000000	2.32830000000000
0.860000000000000	2.30110000000000
0.870000000000000	2.27500000000000
0.880000000000000	2.24900000000000
0.890000000000000	2.22390000000000
0.900000000000000	2.19910000000000
0.910000000000000	2.17520000000000
0.920000000000000	2.15140000000000
0.930000000000000	2.12860000000000
0.940000000000000	2.10570000000000
0.950000000000000	2.08380000000000
0.960000000000000	2.06200000000000
0.970000000000000	2.04090000000000
0.980000000000000	2.02010000000000
0.990000000000000	1.99970000000000
1	1.97990000000000
  };
\addlegendentry{Coded (LQ)}

\addplot [color=black, densely dashed, line width=0.8pt]
  table[row sep=newline]{
 0.500000000000000	7.68790000000000
0.510000000000000	7.53720000000000
0.520000000000000	7.39230000000000
0.530000000000000	7.25280000000000
0.540000000000000	7.11850000000000
0.550000000000000	6.98910000000000
0.560000000000000	6.86430000000000
0.570000000000000	6.74380000000000
0.580000000000000	6.62750000000000
0.590000000000000	6.51520000000000
0.600000000000000	6.40670000000000
0.610000000000000	6.30160000000000
0.620000000000000	6.20000000000000
0.630000000000000	6.10150000000000
0.640000000000000	6.00630000000000
0.650000000000000	5.91380000000000
0.660000000000000	5.82420000000000
0.670000000000000	5.73740000000000
0.680000000000000	5.65290000000000
0.690000000000000	5.57100000000000
0.700000000000000	5.49140000000000
0.710000000000000	5.41410000000000
0.720000000000000	5.33900000000000
0.730000000000000	5.26580000000000
0.740000000000000	5.19460000000000
0.750000000000000	5.12530000000000
0.760000000000000	5.05790000000000
0.770000000000000	4.99220000000000
0.780000000000000	4.92820000000000
0.790000000000000	4.86600000000000
0.800000000000000	4.80620000000000
0.810000000000000	4.74580000000000
0.820000000000000	4.68780000000000
0.830000000000000	4.63130000000000
0.840000000000000	4.57620000000000
0.850000000000000	4.52230000000000
0.860000000000000	4.46980000000000
0.870000000000000	4.41850000000000
0.880000000000000	4.36820000000000
0.890000000000000	4.31910000000000
0.900000000000000	4.27120000000000
0.910000000000000	4.22420000000000
0.920000000000000	4.17820000000000
0.930000000000000	4.13340000000000
0.940000000000000	4.08940000000000
0.950000000000000	4.04630000000000
0.960000000000000	4.00420000000000
0.970000000000000	3.96290000000000
0.980000000000000	3.92240000000000
0.990000000000000	3.88300000000000
1	3.84400000000000
};
\addlegendentry{Coded (HQ)}

\end{axis}
\end{tikzpicture}
        \caption{Error-free, $\mu_1=1$, $\mu_2=1.5$.}
        \label{fig:paoi_eta_unb}
    \end{subfigure}	
    \begin{subfigure}[b]{.49\linewidth}
        \centering
        \begin{tikzpicture}
\pgfplotsset{every tick label/.append style={font=\scriptsize}}
\tikzstyle{dotted}= [dash pattern=on \pgflinewidth off 0.5mm] 
\tikzstyle{dashed}= [dash pattern=on 7.5*0.8*0.8pt off 7.5*0.4*0.8pt]
\tikzstyle{dashdotted} = [dash pattern=on 7.5*0.8*0.6pt off 7.5*0.8*0.3pt on \the\pgflinewidth off 7.5*0.8*0.3pt]
\tikzstyle{dotted2} = [dash pattern=on 7.5*0.8*0.3pt off 7.5*0.8*0.2pt]

\begin{axis}[%
width=\sfwidth,
height=\sfheight,
at={(0,0)},
scale only axis,
xmin=0.5,
xmax=1,
xlabel near ticks,
xlabel style={font=\footnotesize\color{white!15!black}},
xlabel={$\eta$},
ylabel near ticks,
ymin=0,
ymax=15,
ylabel style={font=\footnotesize\color{white!15!black}},
ylabel={$\Delta_{99}$},
axis background/.style={fill=white},
xmajorgrids,
ymajorgrids,
legend style={font=\tiny, at={(0.99,0.98)}, anchor=north east,legend columns=2, legend cell align=left, align=left, fill opacity=0.8, draw opacity=1, text opacity=1, draw=white!80!black}
]

\addplot [color=red, only marks, mark=square, mark size=1.5]
 table[row sep=newline]{
1 5.5084
};
\addlegendentry{Alternating (sim.)}

\addplot [color=cyan, only marks, mark=triangle*, mark size=1.5]
 table[row sep=newline]{
1 5.1074
};
\addlegendentry{Alternating (bound)}

\addplot [color=violet, only marks, mark=o, mark size=1.5]
 table[row sep=newline]{
0.5 6.3784 
};
\addlegendentry{Replicated }

\addplot [color=orange_D, only marks, mark=x, mark size=2]
 table[row sep=newline]{
1 5.5559
};
\addlegendentry{Split }

\addplot [color=red, densely dotted, line width=0.8pt]
 table[row sep=newline]{
0.500000000000000	6.37840000000000
0.510000000000000	6.25350000000000
0.520000000000000	6.13340000000000
0.530000000000000	6.01790000000000
0.540000000000000	5.90660000000000
0.550000000000000	5.79920000000000
0.560000000000000	5.69580000000000
0.570000000000000	5.59610000000000
0.580000000000000	5.49970000000000
0.590000000000000	5.40670000000000
0.600000000000000	5.31660000000000
0.610000000000000	5.22960000000000
0.620000000000000	5.14550000000000
0.630000000000000	5.06380000000000
0.640000000000000	4.98480000000000
0.650000000000000	4.90840000000000
0.660000000000000	4.83410000000000
0.670000000000000	4.76200000000000
0.680000000000000	4.69200000000000
0.690000000000000	4.62410000000000
0.700000000000000	4.55820000000000
0.710000000000000	4.49410000000000
0.720000000000000	4.43180000000000
0.730000000000000	4.37120000000000
0.740000000000000	4.31220000000000
0.750000000000000	4.25490000000000
0.760000000000000	4.19900000000000
0.770000000000000	4.14470000000000
0.780000000000000	4.09180000000000
0.790000000000000	4.03990000000000
0.800000000000000	3.98940000000000
0.810000000000000	3.94030000000000
0.820000000000000	3.89240000000000
0.830000000000000	3.84570000000000
0.840000000000000	3.80030000000000
0.850000000000000	3.75560000000000
0.860000000000000	3.71220000000000
0.870000000000000	3.66950000000000
0.880000000000000	3.62740000000000
0.890000000000000	3.58700000000000
0.900000000000000	3.54710000000000
0.910000000000000	3.50810000000000
0.920000000000000	3.47040000000000
0.930000000000000	3.43280000000000
0.940000000000000	3.39660000000000
0.950000000000000	3.36080000000000
0.960000000000000	3.32590000000000
0.970000000000000	3.29180000000000
0.980000000000000	3.25810000000000
0.990000000000000	3.22550000000000
1	3.19310000000000
 };
\addlegendentry{Coded (LQ)}

\addplot [color=black, densely dashed, line width=0.8pt]
 table[row sep=newline]{
0.500000000000000	11.1115000000000
0.510000000000000	10.8932000000000
0.520000000000000	10.6837000000000
0.530000000000000	10.4826000000000
0.540000000000000	10.2880000000000
0.550000000000000	10.1016000000000
0.560000000000000	9.92060000000000
0.570000000000000	9.74680000000000
0.580000000000000	9.57880000000000
0.590000000000000	9.41610000000000
0.600000000000000	9.25930000000000
0.610000000000000	9.10790000000000
0.620000000000000	8.96100000000000
0.630000000000000	8.81850000000000
0.640000000000000	8.68060000000000
0.650000000000000	8.54700000000000
0.660000000000000	8.41750000000000
0.670000000000000	8.29180000000000
0.680000000000000	8.16990000000000
0.690000000000000	8.05150000000000
0.700000000000000	7.93660000000000
0.710000000000000	7.82510000000000
0.720000000000000	7.71680000000000
0.730000000000000	7.61330000000000
0.740000000000000	7.51010000000000
0.750000000000000	7.40980000000000
0.760000000000000	7.31010000000000
0.770000000000000	7.21580000000000
0.780000000000000	7.12270000000000
0.790000000000000	7.03230000000000
0.800000000000000	6.94480000000000
0.810000000000000	6.85910000000000
0.820000000000000	6.77500000000000
0.830000000000000	6.69390000000000
0.840000000000000	6.61400000000000
0.850000000000000	6.53590000000000
0.860000000000000	6.46080000000000
0.870000000000000	6.38570000000000
0.880000000000000	6.31350000000000
0.890000000000000	6.24240000000000
0.900000000000000	6.17290000000000
0.910000000000000	6.10540000000000
0.920000000000000	6.03870000000000
0.930000000000000	5.97430000000000
0.940000000000000	5.91020000000000
0.950000000000000	5.84860000000000
0.960000000000000	5.78700000000000
0.970000000000000	5.72790000000000
0.980000000000000	5.66900000000000
0.990000000000000	5.61200000000000
1	5.55590000000000
};
\addlegendentry{Coded (HQ)}

\end{axis}
\end{tikzpicture}
        \caption{$\mu_1=\mu_2=1$, $\varepsilon_1=0.1$, and $\varepsilon_2=0.2$.}
        \label{fig:paoi_eta_unberr}
    \end{subfigure}	
     \caption{99th percentile $\Delta_{99}$ of the \gls{paoi} with optimized frame rate as a function of the coding rate $\eta$ in a $D/M/2$.}
 \label{fig:paoi_eta}
\end{figure}

As we did for the system time, we can then analyze the effect of the coding rate on the 99th percentile of the \gls{paoi} in the coded scheme, as shown in Fig.~\ref{fig:paoi_eta}. In this case, we have used an optimized frame rate, i.e., set the inter-frame period that minimizes the 99th percentile of the \gls{paoi} for each value of $\eta$. The two plots on the left, whose scenarios are error-free, show a smaller effect of the coding rate, while it becomes more important for error-prone system. The intuitive understanding that we got from the system time results holds: the more stressed a system is, the more an efficient code matters, and the tighter the trade-off between quality and latency or age becomes. However, while latency-oriented systems are very sensitive to load, \gls{paoi} is very sensitive to the packet erasure probability, as missing a frame can significantly increase the age. The results for the queue-based scheduler are also shown for the two error-free scenarios: in this case, the alternating scheme performs almost as well even in the unbalanced scenario. This is because sending multiple consecutive packets over the same path might improve the average age, but has a negative effect in the worst case, as one hold-up can block two consecutive packets, increasing the age significantly.

\section{Conclusions and future work}\label{sec:concl}

In this work, we analyzed some schemes for the transmission of data over parallel queuing systems, with multipath \gls{vr} as a motivating scenario. Unlike previous works on the fork-join model, we derived the full distribution of the latency and \gls{paoi} for uncoded and coded schemes in the presence of communication errors, and examined the performance of the various schemes as a function of the frame rate and the efficiency of the video encoding. The trade-offs between picture quality, frame rate, and timeliness are complex, and our analysis provides a handy tool for system designers.

The analysis opens several avenues of future research, most of which aim at making the model more realistic. The use of more complex communication models and scheduling schemes is one, while another is a closer examination of real omnidirectional video codecs and traffic models. These two expansions of the work can be combined to reach a realistic scheduling framework, which should also take into account the human perception of smoothness and delay in \gls{vr}.

\section*{Acknowledgment}
This work is part of the IntellIoT project that received funding from the European Union's Horizon 2020 research and innovation program under grant agreement No. 957218.

\bibliographystyle{IEEEtran}
\bibliography{parallel.bib}

\end{document}